\documentclass[journal,twoside,final]{IEEEtran}
\usepackage{enumitem}
\usepackage{amsfonts}
\usepackage{amsmath}
\usepackage{bm} 
\usepackage{caption}
\usepackage{subcaption}
\usepackage{url}
\usepackage{color,colortbl}	
\usepackage[dvipsnames]{xcolor}
\usepackage{soul}
\usepackage[colorlinks]{hyperref}
\hypersetup{
    linkcolor=blue,
    filecolor=magenta,
    urlcolor=teal,
    citecolor=magenta
    }
\usepackage{tikz}
\usetikzlibrary{arrows}
\usepackage{arydshln}
\usepackage{mathtools}

\begin{document}

\title{Electromechanical Dynamics of the Heart: A Study of Cardiac Hysteresis During Physical Stress Test}

\author{Sajjad~Karimi, Shirin~Karimi, Amit~J~Shah, Gari~D.~Clifford~\IEEEmembership{Fellow,~IEEE}, and Reza~Sameni\textsuperscript{*} \IEEEmembership{Senior~Member,~IEEE}%
\thanks{Sajjad~Karimi, A.J.~Shah, G.D.~Clifford and R~Sameni are with the School of Medicine, Emory University. Shirin~Karimi is an independent researcher in Atlanta, GA. A.J. Shah is also with the Rollins School of Public Health, Department of Epidemiology, Emory University. G.D.~Clifford and R.~Sameni are also with the Department of Biomedical Engineering, Georgia Institute of Technology and Emory University (email: \url{rsameni@dbmi.emory.edu}).
}%
}
\maketitle
\begin{abstract}
Cardiovascular diseases are best diagnosed using multiple modalities that assess both the heart’s electrical and mechanical functions. While imaging techniques like echocardiography and nuclear imaging are most effective, they are costly and not widely accessible. More affordable and accessible technologies, such as electrocardiography (ECG) and phonocardiography (PCG), when recorded simultaneously, may provide valuable insights into the heart’s electromechanical coupling and could be useful for prescreening in low-resource settings.

Using physical stress test data from the EPHNOGRAM simultaneous ECG-PCG dataset, collected from 23 healthy male subjects (age: 25.4$\pm$1.9 yrs), we investigate electromechanical intervals (RR, QT, systolic, and diastolic) and their interactions during exercise, along with the phenomenon of \textit{hysteresis} between cardiac electrical activity and mechanical responses.

Time delay analysis of electromechanical intervals reveals distinct temporal relationships between the QT, systolic, and diastolic intervals, with the RR interval as the primary driver. The diastolic interval shows near-synchrony with RR intervals, while the QT interval responds to RR interval changes with an average delay of 10.5\,s. The systolic interval, however, displays a slower response, with an average delay of 28.3\,s. Along with QT-RR hysteresis, we examined systolic-RR and diastolic-RR hysteresis, finding narrower hysteresis loops for diastolic RR and wider loops for systolic RR. Significant correlations (average 0.75) were found between heart rate changes and hysteresis loop areas for QT, systolic, and diastolic intervals. We propose the equivalent circular area diameter as a promising biomarker for cardiac function during physical exercise stress tests.

Finally, deep learning models, including Long Short-Term Memory (LSTM) and Convolutional Neural Networks (CNN), were used to estimate the QT, systolic, and diastolic intervals from the RR interval. Root mean square error analysis showed that CNN and LSTM models outperformed simpler memoryless linear and neural networks. This confirms the nonlinear dynamic relationship between the RR interval and other cardiac intervals, as evidenced by hysteresis loops.

These findings suggest a significant cardiac \textit{memory effect}, reflected in the ECG and PCG morphology and timing, and linked to heart rate history.

\end{abstract}

\section{Introduction}
\label{sec:introduction}
Cardiovascular diseases (CVDs) are diverse and most accurately identified through a sequence of clinical examinations, including auscultation, electrocardiography (ECG), echocardiography, and cardiac imaging. Although each technique has the capacity to identify specific aspects of CVDs, there is an increasing bias and reliance on cardiac imaging. Conditions such as ischemic heart disease, heart failure, and valvular heart disease are frequently diagnosed using advanced imaging modalities like echocardiography and nuclear imaging. While these techniques are highly effective, their high cost limits accessibility in low-resource settings. Additionally, they are impractical for ambulatory monitoring, which requires prolonged patient observation during daily activities.

Combining multiple low-cost monitoring technologies such as ECG and phonocardiography (PPG)~\cite{Oliveira2022,Reyna2023}, in a portable or wearable device is a promising solution. Complementary modalities, especially when acquired and processed simultaneously (time-synchronously), can provide a more holistic perspective of the electromechanical function of the heart~\cite{Kazemnejad2024,EPHNOGRAMDataset}.

Electromechanical multimodality can be specifically used to analyze the mutual effects and interactions between the characteristics of different modalities. For instance, the impact of exercise, stress, and residual effects on the electrical and mechanical functions of the heart can be better investigated through multimodality approaches. While clinical ECG and cardiac stress tests are commonly studied as static ``snapshots,'' previous research has reported hysteresis-like effects---implying short- and long-term memory---where the characteristics of cardiac modalities such as ECG depend on the patient's longer-term condition. This includes whether the heart is transitioning from relaxation to higher activity or moving towards higher intensity activities. QT-RR hysteresis indicates that, for a given R-R interval, the QT interval duration is different (typically shorter) during recovery after exercise than during the exercise itself~\cite{LAUER2006315}.

The hysteresis effect has been previously reported in QT analysis vs heart rate~\cite{fossa2005dynamic}, showing that the QT interval is not only a function of the heart rate, but also dependent on the heart rate history. This implies a non-static relationship between the QT interval and the heart rate and necessitates the simultaneous study of the heart rate trend (beyond the current heart rate) in studying the QT or the corrected QT (QTc) in stress-test (non resting condition) scenarios. QT-RR hysteresis has been proposed as a robust and potentially independent predictor of myocardial ischemia, offering insights beyond ST-segment measures~\cite{LAUER2006315}. It is considered a strong indicator of reduced myocardial perfusion and may aid in the noninvasive assessment of early coronary artery disease by reflecting mechanical and biochemical changes~\cite{STAROBIN2007S91}. While the clinical utility of QT-hysteresis evaluation shows promise, further studies are needed to validate its relevance in specific conditions, with a focus on methods that separately estimate QT-hysteresis and QT-RR dependency~\cite{GravelClinical}. Other studies have investigated the relationships between blood pressure, cardiac vagal baroreflex function, and heart rate, revealing similar hysteresis effects between other cardiac functions~\cite{studinger2007mechanical}.

The investigation of pulse transit time (PTT) and systolic blood pressure (SBP) indicates the hysteresis relationship and reveals insights into autonomic nervous system function~\cite{liu2013attenuation}. The relationship between SBP and PTT is influenced by the mechanical properties of the arterial wall, primarily modulated by the sympathetic nervous system's effects on vascular smooth muscle (VSM) tone during exercise. 
In the context of postexercise hypotension and changes in autonomic function, baroreflex sensitivity and hysteresis are investigated to explore how the baroreflex responds differently to rising and falling blood pressure, hypothesizing that postexercise, the relative decrease in falling blood pressure would exacerbate hysteresis~\cite{studinger2007mechanical, willie2011neuromechanical}. Despite these efforts, to date, systolic-RR hysteresis between the mechanical (via PCG) and electrical (via ECG) behaviors of the heart has not been studied. Hypothetically, when accurately quantified, cardiac hysteresis can serve as a biomarker for cardiovascular health assessment.



In this research, we use exercise stress test recordings from the EPHNOGRAM simultaneous ECG-PCG dataset to investigate and analyze electromechanical intervals, including RR, QT, systolic, and diastolic intervals \cite{Kazemnejad2024,EPHNOGRAMDataset}. The contributions of the current study are summarized as follows:
\begin{enumerate}
    \item \textit{Time delay analysis:} We quantify the response speed of QT, systolic, and diastolic intervals to changes in RR intervals during stress tests.
    \item \textit{Hysteresis patterns:} Beyond QT-RR hysteresis, we identify systolic-RR and diastolic-RR hysteresis based on multimodal cardiac data processing. These hysteresis patterns reveal dynamic relationships between intervals during stress and recovery.
    \item \textit{Nonlinear dynamic:} Deep learning models, combined with RMSE analysis, validate the presence of a nonlinear dynamic relationship between the RR interval and other cardiac intervals.
    \item \textit{A novel marker for cardiac hysteresis analysis:} We explore a strong correlation between heart rate changes and the equivalent circular area diameter ($D_a$) of hysteresis loops. This correlation exists across all subjects and is valid for QT, systolic, and diastolic intervals
\end{enumerate}


The remainder of this paper is structured as follows: In Section~\ref{sec:hysteresis}, we review the notion of hysteresis and previous literature on cardiac hysteresis. Section~\ref{sec:dataset} details the EPHNOGRAM cardiac stress-test dataset. In Section~\ref{sec:method}, we present the processing steps to transform raw ECG and PCG signals into informative electromechanical time intervals, detailing algorithms for fiducial point extraction, signal segmentation, interval extraction, and synchronization. Section~\ref{sec:result} presents the results, analyzing ECG morphology changes, time delay relationships between RR intervals and other cardiac intervals (QT, systolic, diastolic), and observed hysteresis characteristics, supported by visualizations and statistical analyses. Section~\ref{sec:discussion} discusses the results and their implications, including the physiological and clinical relevance of observed patterns, study limitations, and future research directions. The final section, Section~\ref{sec:conclusion}, summarizes the key contributions and significance of this work in understanding cardiac function during stress tests.

\section{Hysteresis}
\label{sec:hysteresis}
In physics and engineering, hysteresis refers to the phenomenon where the output of a system depends not only on its current input but also on its past inputs. This is typically seen in systems with magnetic materials, where the magnetic flux density ($B$) lags behind the magnetic field strength ($H$). This creates a looped graph, known as the \textit{hysteresis loop} when plotting $B$ against $H$. The area within this loop represents energy loss due to the lagging response. Hysteresis is important in applications such as transformers and inductors, where it affects the efficiency and behavior of the devices.

In the ECG context, hysteresis effects have been observed when the QT interval's duration lags behind changes in heart rate (RR interval), and this lag is a function of the current heart rate and its previous trend~\cite{fossa2005dynamic}. Similar to magnetic hysteresis, where a material's magnetic properties lag behind the force that creates them, QT hysteresis indicates how the QT interval responds to changes in the heart rate. This lag can vary depending on the rate of change in heart rate; slower changes may reduce the lag, while faster changes may increase it~\cite{martin2022qt}.
The QT vs RR plot shows that the trajectory is not a linear/nonlinear curve. Instead, depending on whether the heart rate is increasing or decreasing, a different path is taken in the QT-RR phase plane. This indicates that changes in the RR interval do not immediately translate to changes in the QT interval, resulting in elliptic trajectories in the QT-RR plane. The practical implication of the hysteresis effect is that in non-resting conditions, merely reporting the QT interval or the corrected QT interval (QTc) is inadequate; in stress-test QT intervals (and other electromechanical measurements from the heart, as shown in this work) depend on whether the heart rate is accelerating or decelerating.  


The QT hysteresis is considered an arrhythmic risk marker~\cite{gravel2018clinical}. Recent studies have focused on evaluating this lag during stress tests, observing a reduction as HR increases, though the mechanisms remain unclear. Factors like the autonomic nervous system (ANS), particularly sympathetic activity, may influence this adaptation. Computational models suggest that $\beta$-adrenergic stimulation, especially nearing peak stress, may play a role in reducing the QT lag in tracking the heart rate. Identifying these patterns could aid in understanding arrhythmic risk during stress tests~\cite{perez2023role}.

Studies have shown that athletes may exhibit physiological QT prolongation due to chronic exercise training~\cite{christou2022prolonged}. This poses challenges in distinguishing between normal adaptation and potential pathology like long QT syndrome (LQTS). Diagnostic evaluations, including exercise testing, play a crucial role in identifying athletes at risk, especially in borderline cases of QT prolongation~\cite{christou2022prolonged}. It has been shown that the relationship between QT and RR intervals is highly individual-specific~\cite{batchvarov2002individual}, likely due to the dynamic nature of the system, subjects' physical preparedness, and individual anatomical and physiological variations~\cite{malik2002relation}.

\section{The EPHNOGRAM simultaneous ECG-PCG dataset}
\label{sec:dataset}
The EPHNOGRAM dataset, publicly available on PhysioNet, is used for this study \cite{Kazemnejad2024,EPHNOGRAMDataset}. It contains simultaneous ECG and PCG data collected from 24 healthy adult participants during various physical activities, including resting, walking, running, and stationary biking, recorded in an indoor fitness center. The EPHNOGRAM project aimed to enhance the understanding of cardiac function and explore the interrelationships between the heart's mechanical and electrical activities. The dataset includes data from a total of 24 male subjects aged between 23 and 29 years (average: 25.4$\pm$1.9 years). The dataset consists of 68 simultaneous ECG and PCG recordings, including 8 recordings of 30-second duration and 60 recordings of 30-minute duration, acquired from a single-channel ECG and a single PCG stethoscope. Each volunteer performed the following physical activities \cite{Kazemnejad2024}:



\begin{enumerate}
    \item \textit{Scenario A --- resting condition:} Participants lay horizontally on a bed in a quiet room for 30-minute ECG and PCG recordings. Additionally, a few 30-second samples were recorded while participants sat in an armchair.
   
    \item \textit{Scenario B --- walking condition:} The participant walked on a treadmill at a constant speed of 3.7\,km/h.
    
    \item \textit{Scenario C --- treadmill stress-test:} The stress test utilized the modified Bruce protocol~\cite{bruce1971exercise}, lasting 30 minutes. Speed and incline increased until subjects reached fatigue, excessive heart rate, or chest pain. At this point, speed was gradually decreased to 3\,km/h and the treadmill was inclined horizontally. Participants walked briefly before stopping and sitting until the test ended, with signals acquired throughout.
    \item \textit{Scenario D --- bicycle stress-test:} 
    The bicycle stress-test protocol lasted 30 minutes per session. Participants began with a 2-minute rest on a stationary bike before pedaling at an increasing workload of 25\,Watts per minute. They adjusted power consumption to match specified levels, though maintaining a constant speed was challenging. Testing ceased upon excessive fatigue, heart rate, or chest pain, with workload gradually reduced to 25\,Watts per minute. Participants rested on the bike until the session ended, with signals continuously recorded by the EPCG device.
\end{enumerate}

Initially, ten subjects participated. Five did not continue to the next stages. Six new volunteers joined, making a total of eleven in each of the three subsequent scenarios. In Scenario D, two volunteers could not complete the bicycle exercise stress test due to physical fatigue. All records are in the PhysioNet WFDB format and include annotations regarding the physical activity intensity. Further details regarding the data collection protocol can be followed from~\cite{Kazemnejad2024,EPHNOGRAMDataset}.


\subsection{Selected stress test recordings}
The current study focuses on stress testing. We therefore selected records from scenarios C and D involving treadmill and bicycle exercises. After screening using the \texttt{ECGPCGSpreadsheet.csv} spreadsheet available in \cite{EPHNOGRAMDataset}, some recordings were deemed extremely noisy or disconnected and were excluded from our analysis. Ultimately, 23 stress test recordings meeting cleanliness criteria and completed by subjects were chosen for analysis. In \cite{EPHNOGRAMDataset}, the data files corresponding to these recordings are labeled \texttt{ECGPCG00XY.mat}. We selected the records: \texttt{XY} = \texttt{01}, \texttt{17}, \texttt{18}, \texttt{25}, \texttt{26}, \texttt{27}, \texttt{29}, \texttt{30}, \texttt{32}, \texttt{33}, \texttt{34}, \texttt{36}, \texttt{37}, \texttt{38}, \texttt{47}, \texttt{52}, \texttt{55}, \texttt{61}, \texttt{62}, \texttt{64}, \texttt{66}, \texttt{67}, and \texttt{68}.

\section{Method}
\label{sec:method}
\begin{figure*}[tb]
    \centering
    \includegraphics[trim=0in 0in 0in 0in, clip, width=0.99\linewidth]{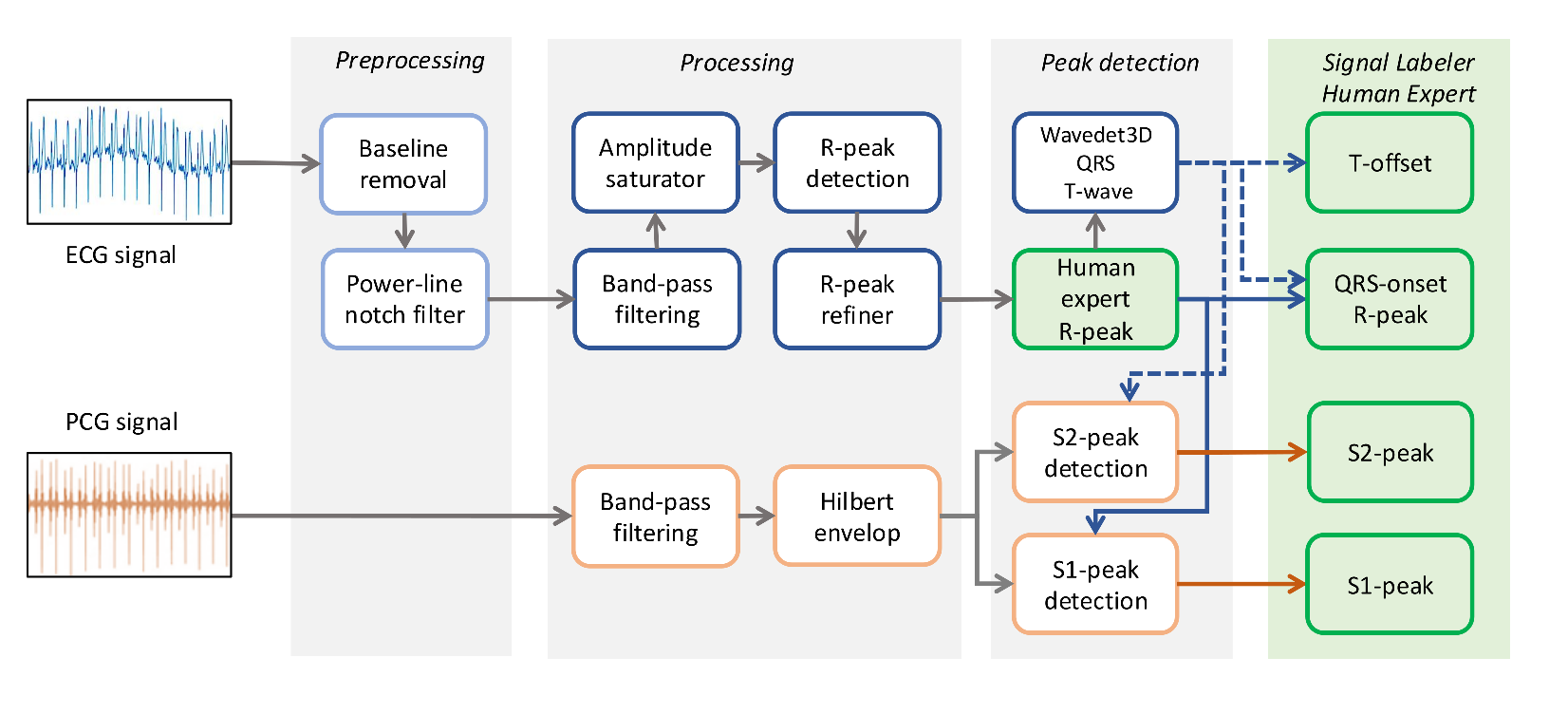}
    \caption{Signal processing block diagram for extracting fiducial points from ECG and PCG data}
    \label{fig:processing}
\end{figure*}

The primary objective of this study is to explore the time intervals between ECG and PCG components and to study the morphological variations of these modalities under stress test. Various ECG- and PCG-based fiducial points are commonly used in the literature. Among the various intervals derivable from ECG and PCG signals, we focus on four key intervals renowned for their clinical significance and prevalence in the literature: the RR and QT intervals from the ECG signal, and the systolic and diastolic intervals from the PCG signal. The initial step involves identifying fiducial points within the ECG and PCG, such as QRS onset, R peak, and T-wave offset from the ECG, and identifying S1 and S2 peaks from the PCG power envelope. For this, the processing steps depicted in Fig.~\ref{fig:processing} were implemented on both the ECG and PCG channels of the records. The outcomes of this processing framework include the fiducial points from both ECG and PCG, as well as beat-wise annotations for each time-series. A detailed breakdown of each stage within the processing unit is provided below.

\subsection{Preprocessing}
\label{sec:preprocessing}
In the ECG channels, we estimated the baseline wander by applying a moving median filter with a sliding window length of 0.6\,s, followed by a moving average filter with a sliding window length of 0.3\,s. The sliding windows were applied in a non-causal manner, to avoid adding any time delays between the input and output, therefore maintaining the synchrony between the ECG and PCG channels.

The resulting baseline estimate was subtracted from the ECG. This two-stage combination has proven highly effective and robust for ECG baseline wander removal~\cite{SSJ08,jamshidian2018fetal}. Unlike ECG, PCG as an acoustic signal does not inherently possess baseline drift. Therefore, any non-zero average on PCG is typically attributed to electronic circuitry and can be safely eliminated with a DC blocker, without affecting the PCG's frequency content. 

Given the indoor data acquisition setting, the presence of power-line noise (at 50 Hz in this dataset) was inevitable, manifesting with variable amplitudes throughout the 30-minute acquisition sessions, sometimes as brief burst noises. To address this, a fixed second-order IIR notch filter with a Q-factor of 45 was employed in the current study to attenuate the power-line noise. The filter was applied using forward-backward filtering (the \texttt{filtfilt} function in MATLAB and Python's \texttt{scipy.signal} package).


\subsection{Fiducial point extraction}

\subsubsection{ECG-based features}
\label{sec:ECGheartrate}
We implemented an ECG R-peak detector inspired by the Pan-Tompkins algorithm~\cite{Pan1985}. Initially, a bandpass FIR filter with a passband ranging from 10\,Hz to 40\,Hz was applied to the ECG signal. Subsequently, the amplitude of the filtered signal was saturated using a hyperbolic tangent function $y = \alpha\tanh(x/\alpha)$, with $\alpha$ set to $k$ times the standard deviation of the bandpass filtered signal ($k=10$ for the results reported later). The saturation step ensures that short-term burst/spike noises and motion artifacts do not affect subsequent peak detection and thresholding steps of the algorithm. Then, the power envelope of the filtered signal was computed using a sliding window of length 75\,ms. Finally, R-peaks were identified through local peak detection of the power envelopes within heart rate adaptive window lengths. The robust R-peak detector, \texttt{peak\_det\_likelihood\_long\_recs.m}, which is specifically adapted for long records, is available in the open-source electrophysiological toolbox (OSET) \cite{OSET3.14}.

In addition to the R-peaks an accurate fiducial detection algorithm was required to identify the QRS onset and T-wave offset for computing QT intervals per ECG beat. To accomplish this, a fiducial detection algorithm based on the Latent Structure Influence Model (LSIM) was implemented and utilized to detect QRS and T-wave onset and offset~\cite{karimi2020tractableinf, karimi2023tractablemle}. This algorithm, denoted as the LSIM-FD block in Fig.~\ref{fig:processing}, takes the ECG signal and the R peaks as inputs, and returns the QRS and T-wave fiducial points. Source codes for the R peak and our LSIM-based fiducial point detection algorithm are accessible in the Open-Source Electrophysiological Toolbox (OSET)~\cite{OSET3.14}.

We developed a graphical tool using MATLAB's Signal Labeler app to visualize and refine the automatically detected R-peaks. A biomedical engineer specializing in ECG analysis utilized this tool to thoroughly inspect the data and the automatic annotations beat-by-beat, making necessary corrections to any missed or inaccurately detected R-peaks or fiducial points. In a second step, a physician experienced in ECG data labeling overread the annotations performed by the software and the biomedical engineer, to confirm the accuracy of the labels for subsequent analysis. 

\subsubsection{PCG-based features}
\label{sec:PCGheartrate}
PCG-based beat detection and annotation present greater challenges and are less established compared to ECG annotation. One advantage of simultaneous ECG-PCG acquisition and processing is that we can utilize ECG-based R-peaks as references for PCG beat detection and segmentation of its components (S1, S2, etc.), which can be challenging to detect independently~\cite{Kazemnejad2024}. In this study, following the detection of the ECG R-peaks, these peaks were used as initial reference points for estimating the S1 and S2 segments of the PCG, derived from the PCG's power envelope.

\begin{figure}[tb]
    \centering
    \includegraphics[trim={1cm 0cm 1cm .7cm},clip,width=0.99\linewidth]{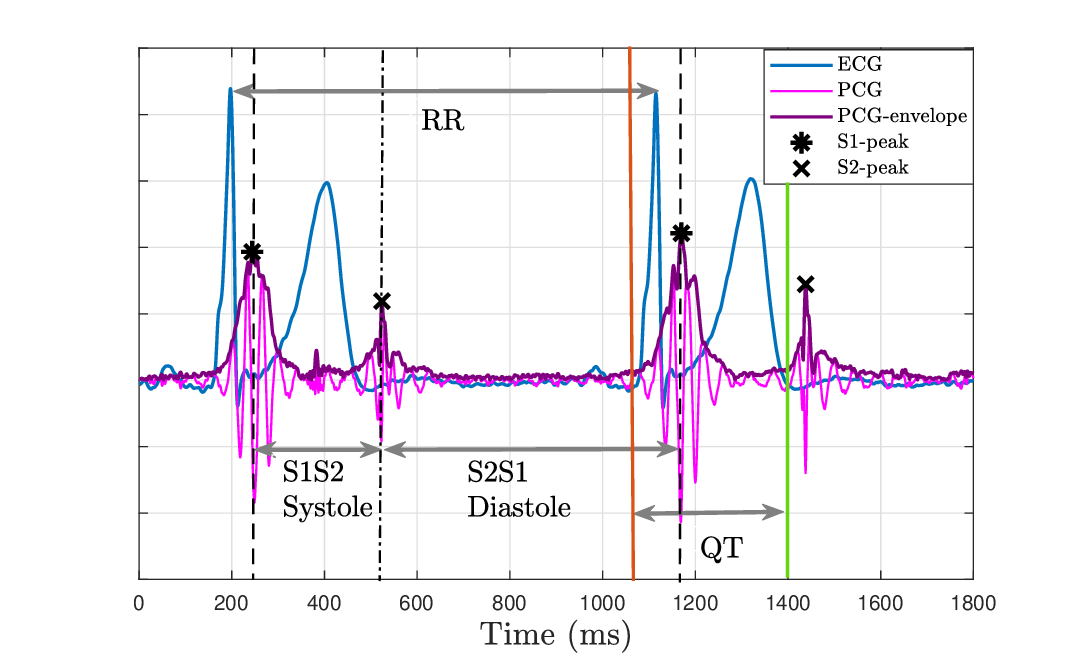}
    \caption{Two cycles of normalized ECG, PCG, and the PCG envelope with the description of considered Electromechanical intervals}
    \label{fig:SimultaneousECGPCG}
\end{figure}

Typical beats of the simultaneous ECG and PCG records from the dataset are shown in Fig.~\ref{fig:SimultaneousECGPCG}. The S1 component is identified as the first dominant PCG peak following the R-peak, while the S2 component is identified as the first dominant PCG peak following the ECG T-wave, consistent with the heart's electrophysiology. To accurately delineate the S1 and S2 waves, estimated ECG fiducial points, such as peaks, onsets, and offsets of key ECG components (including T-waves previously estimated by LSIM-FD in the previous section), were used.

To enhance the visibility and detectability of the S1 and S2 components, the PCG was bandpass filtered with an FIR filter with a passband range of 20.0 to 200.0\,Hz. This choice aligns with previous literature indicating that the dominant frequency components of the PCG fall below 200\,Hz~\cite{akram2018analysis,Kazemnejad2024}. Following this filtering step, the S1 and S2 components of the PCG exhibit distinctive shapes characterized by narrow-band oscillatory waves modulated over bumpy shapes. Consequently, unlike the ECG, relying solely on local peaks of the PCG for component detection would be unreliable. Instead, the Hilbert transform envelope was employed to aid in the detection of the S1 and S2 components. The S1 peak was subsequently identified as the dominant peak of the PCG power envelope occurring between the R-peak and the T-wave peak of the corresponding ECG beat. Similarly, the S2 peak was determined as the dominant peak of the PCG envelope located between the T-wave peak and the subsequent R-peak. After detecting the R-peaks, all S1 and S2 components of the PCG were carefully reviewed by two human annotators (a biomedical engineer and a physician).

\subsection{Multimodal feature set}
The developed signal processing pipeline generates five sets of beat-wise annotations: R-peak indexes, T-wave offset indexes, QRS complex onset indexes, S1-wave peak indexes, and S2-wave peak indexes, as illustrated in Fig.~\ref{fig:processing}.
Having the R-peak and T-offset from the ECG and the S1 and S2 peaks of the PCG, the following time intervals were calculated per beat:
\begin{enumerate}
    \item RR: from one R-peak to the next R-peak,
    \item QT: from the QRS onset to the T-wave offset of the same ECG beat,
    \item S1S2: from one S1-peak to the S2-peak of the same beat (the \textit{systolic} time interval),
    \item S2S1: from one S2-peak to the S1-peak of the next beat (the \textit{diastolic} time interval).
\end{enumerate}
These intervals are illustrated in Fig.~\ref{fig:SimultaneousECGPCG}. When studied as time-series (as a function of beat indexes), electromechanical intervals are inherently non-uniformly sampled data~\cite{cebollada2021mechanisms}. For instance, higher heart rates correspond to shorter RR interval values, causing successive RR samples to be closer together in time. Therefore, to construct a uniformly sampled time-series, we resampled all time-interval features to a uniform sampling rate of 10\,Hz. MATLAB's \texttt{resample} function was used for this interpolation to achieve a uniformly sampled time-series, for further analysis.


\subsection{Equivalent circular area diameter of hysteresis}

Quantifying cardiac hysteresis, including QT-RR hysteresis, is an essential step for its usage as a cardiac biomarker, as it provides insights into the heart's adaptation to both physiological and pathological states~\cite{montull2023hysteresis,montull2020hysteresis}. Yet, modeling systems incorporating hysteresis pose challenges due to their nonlinear nature, requiring advanced mathematical approaches to effectively capture and model their dynamic behavior.

\begin{figure}
    \centering\includegraphics[trim={8cm 2.5cm 11cm 3cm},clip,width=0.6\linewidth]{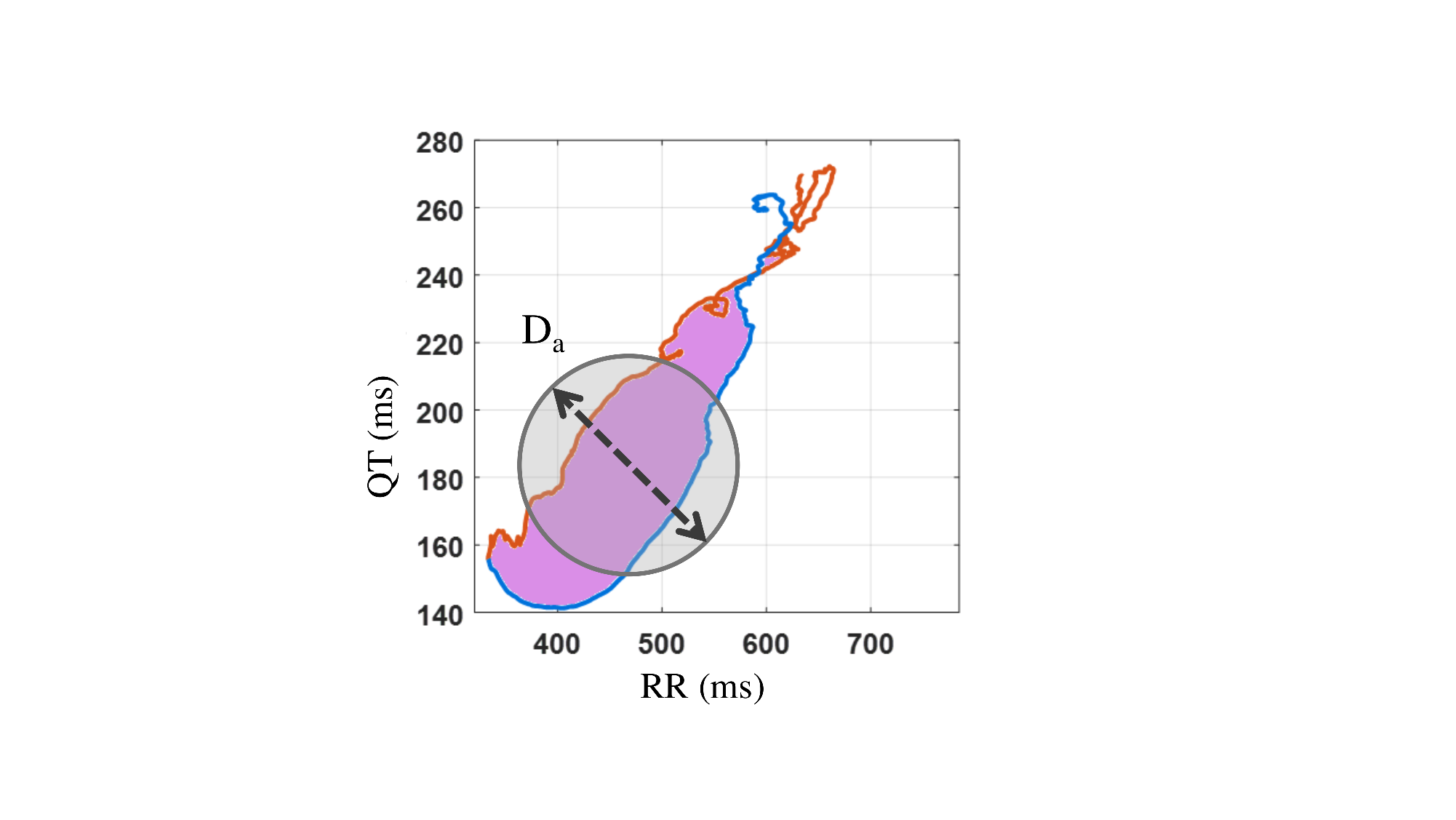}
    \label{fig:ECGbeats-a}
    \caption{Equivalent circular area diameter ($D_a$) for a sample RR-QT hysteresis loop with the equal areas}
\label{fig:LoopDa}
\end{figure}

Hysteresis effects between different variable pairs can have arbitrary shapes. Quantifying these shapes can be challenging due to various factors, such as the type and intensity of the exercise, the duration of data collection, and individual differences like the subject's physical condition and readiness. We propose to use the hysteresis area bounded between their trajectories to quantify the intensity of hysteresis effects. For example, for QT-RR hysteresis loops, narrower loops indicate lower lag, suggesting a more rapid adaptation of the QT interval to changes in heart rate, while wider loops indicate higher lag, suggesting a slower adaptation. Fig.~\ref{fig:LoopDa} shows the hysteresis area between the heart rate acceleration and deceleration phases.

To quantify the area bounded by the hysteresis loops, we propose using the \textit{equivalent circular area diameter} ($D_a$) of hysteresis areas as subjects undergo a full exercise test. Starting at a resting condition with no heart rate acceleration/deceleration, the intensity of the physical activity is gradually increased up to a maximum load, and decreased again gradually back to the resting condition at the same heart rate as the initial point. This full cycle results in bounded (or approximately bounded) hysteresis areas that can be quantified using $D_a$.

The equivalent circular area diameter has been previously used in the context of Particle Size Descriptors~\cite{li2005comparison}. It represents the diameter of a circle/sphere with the same projected area as the hysteresis area, as illustrated in Fig.~\ref{fig:LoopDa}. This parameter, widely utilized in size characterization, can be computed as:
\begin{equation}
D_a = 2\sqrt{\frac{A}{\pi}}      
\end{equation}
where $A$ is the area of the outer loop (\textit{convex hull}) of the two-dimensional scatter plot of the two variables of interest. If both variables share the same unit of measurement, $D_a$ will have the same unit, which for cardiac event time intervals will be in seconds.

\section{Result}
\label{sec:result}
We next delve into a detailed analysis of the dynamic characteristics of various ECG and PCG-based characteristics.

\subsection{ECG and PCG-based parameter variations}
\begin{figure*}[tb]
    \includegraphics[trim={2cm 0cm 0cm .5cm},clip,width=0.99\linewidth]{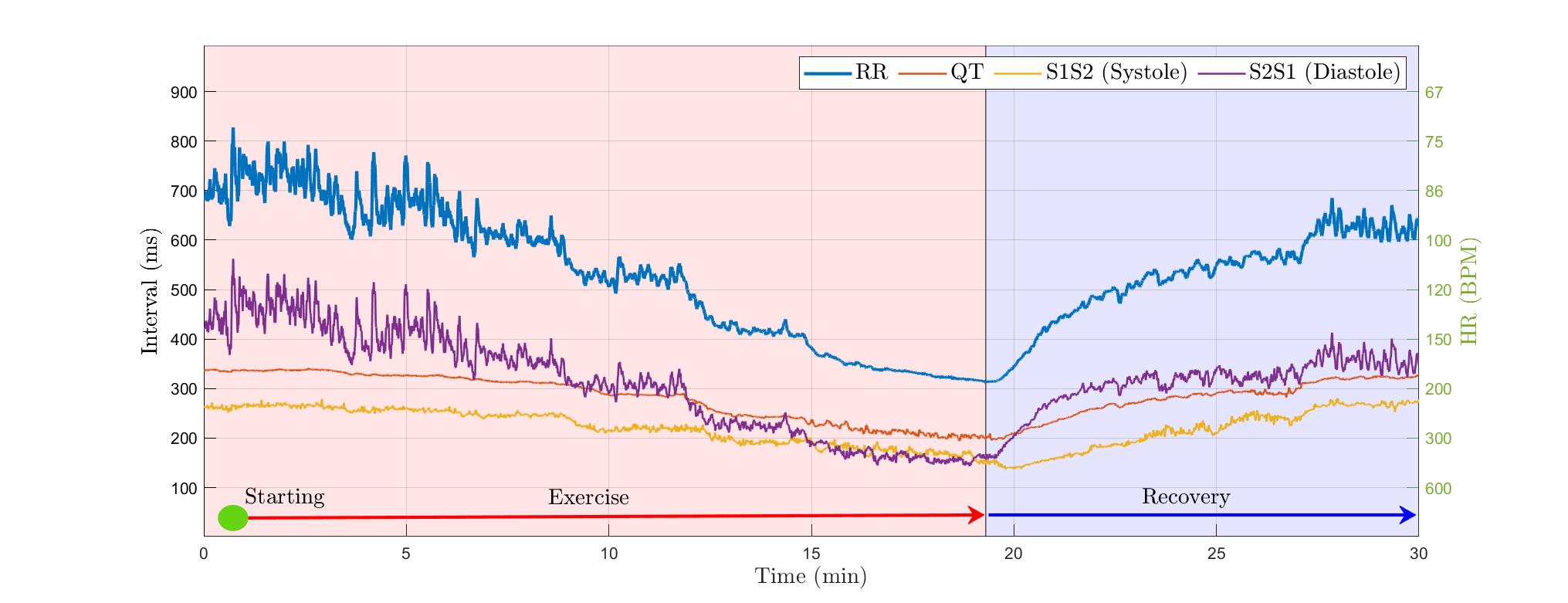}
    \label{fig:PedalingTimeSeries}
\caption{ECG and PCG-based electromechanical intervals including RR, QT, systolic, and diastolic intervals in Bruce treadmill stress test for record \texttt{ECGPCG0038} (exercise: pink; recovery: lavender blue). Arrows show the stress test phases.}
\label{fig:Intervals}
\end{figure*}

Fig.~\ref{fig:Intervals} illustrates the variation and dynamics of the extracted intervals during a Bruce treadmill stress test for record \texttt{ECGPCG0038}, spanning a duration of 30 minutes. The left vertical axis represents the intervals in milliseconds (ms), while the right axis indicates the equivalent heart rate in beats per minute (bpm). As noted in the EPHNOGRAM dataset description \cite{Kazemnejad2024}, the recording comprises an exercise phase marked by increased heart rate (HR), followed by a recovery phase after the subject experiences pressure, excess fatigue, or chest pain. During the recovery phase, the HR returns to normal conditions. The lavender blue and pink backgrounds in Fig.~\ref{fig:Intervals} correspond to the exercise and recovery phases, which are essential to the subsequent analyses.

According to Fig.~\ref{fig:Intervals}, during low-intensity workloads and resting periods, heart rate exhibits greater variability compared to high-intensity exercise phases. This variability is particularly prominent in the S2S1 (diastolic) interval, suggesting that the diastolic period, influenced by the mechanical behavior of the heart, plays a crucial role in adjusting to heart rate variations at different workload levels. 

subsection{ECG/PCG beats waterfall plots}
We now study the morphological changes of the ECG and PCG in varying heart rates. To conduct this analysis, all ECG beats were segmented into 600\,ms windows and aligned based on the R-peak indices. Similarly, the PCG envelopes were extracted based on the ECG R-peak indices to help compare the time synchrony between the two modalities. The segment time windows spanned from 200\,ms before nd 400\,ms after each R-peak for all ECG and PCG beats. This approach allows for a comprehensive examination of the changes in ECG and PCG waveforms as the heart rate fluctuates, providing insights into the underlying physiological processes.

\begin{figure}
    \centering
    \begin{subfigure}{0.95\linewidth}
        \centering
        \includegraphics[trim={0cm 0cm 1cm 0cm},clip,width=\columnwidth]{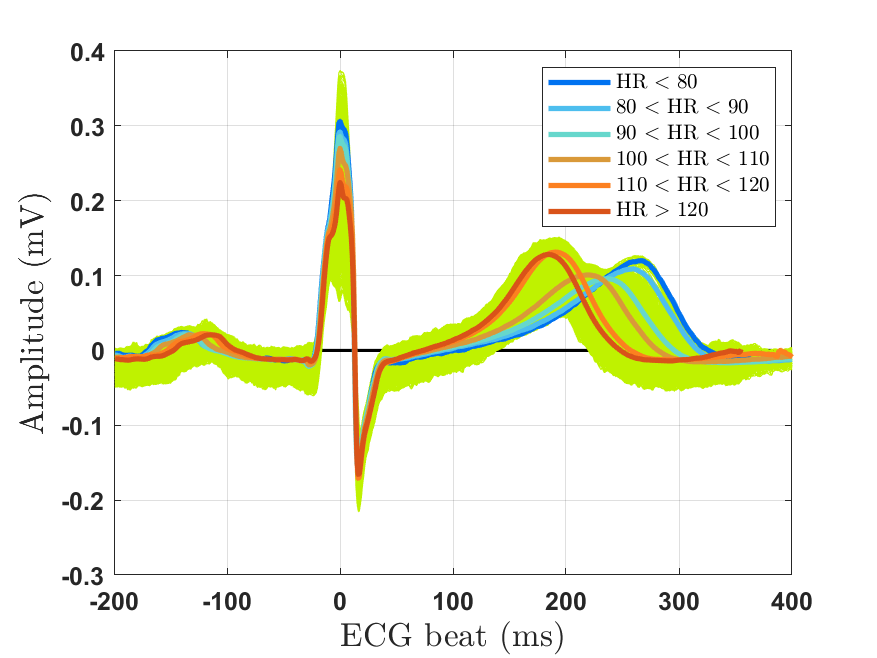}
        \caption{ECG beats}
        \label{fig:ecg_beat_superpositions}
    \end{subfigure}
    \begin{subfigure}{0.95\linewidth}
        \centering
        \includegraphics[trim={0cm 0cm 1cm 0cm},clip,width=\columnwidth]{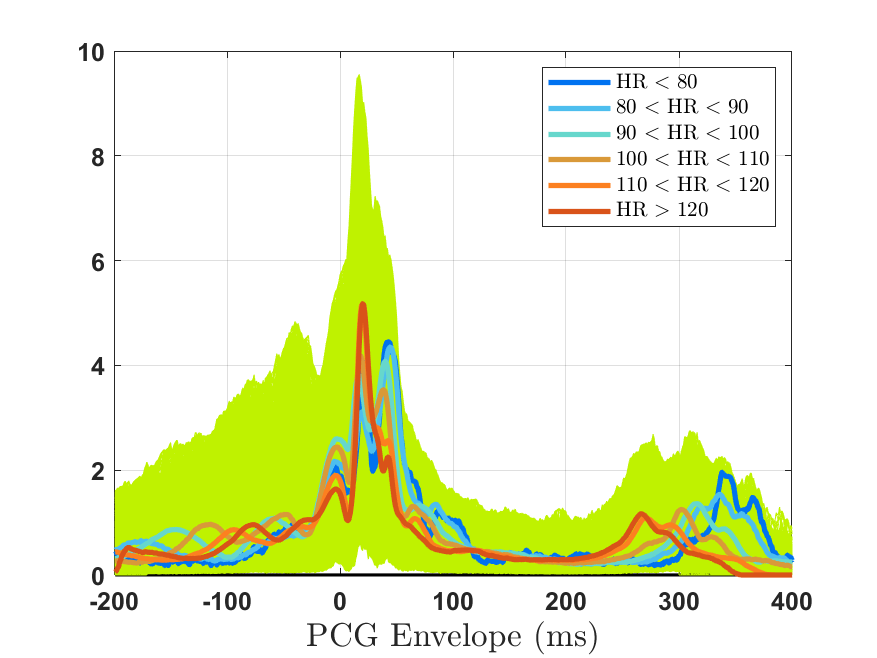}
        \caption{PCG beats}
        \label{fig:pcg_beat_superpositions}
    \end{subfigure}
    \caption{The superposition of all ECG beats and PCG envelopes, and the average beat of clusters for heart rates below 80, 80-90, 90-100, 100-110, 110-120, and above 120 beats per minute. The subject was pedaling on a stationary bicycle (\texttt{ECGPCG0032}).}
\label{fig:ECGbeats}
\end{figure}

As an initial step, Fig.~\ref{fig:ECGbeats} illustrates the superposition of all ECG beats recorded during pedaling on a stationary bicycle. The blue to red colormaps in this figure represent the robust average beat of clusters for heart rates in the ranges below $80$, $[80-90)$, $[90-100)$, $[100-110)$, $[110-120)$, and $120$ or above beats per minute ~\cite{leski2002robust}. Accordingly, the most prominent change in the ECG morphology is in the T-wave segment (i.e., the repolarization cycle), while the QRS complex remains relatively intact across different heart rate clusters. There is also a small change in the P-wave morphology as it exhibits a shift towards the R-peak for ECG beats at higher heart rates. Considering that the R-peaks are used to align the beats in Fig.~\ref{fig:ECGbeats}, the reduced PR interval can be associated with a faster depolarization of the ventricles at higher heart rates. Note that in this figure, the heart rate cluster thresholds are arbitrary and are selected for visualization purposes only. Otherwise, there is a continuous drift in the ECG morphology from low to high heart rates.


\begin{figure*}
    \includegraphics[trim={0cm 1cm 0cm .5cm},clip,width=0.95\linewidth]{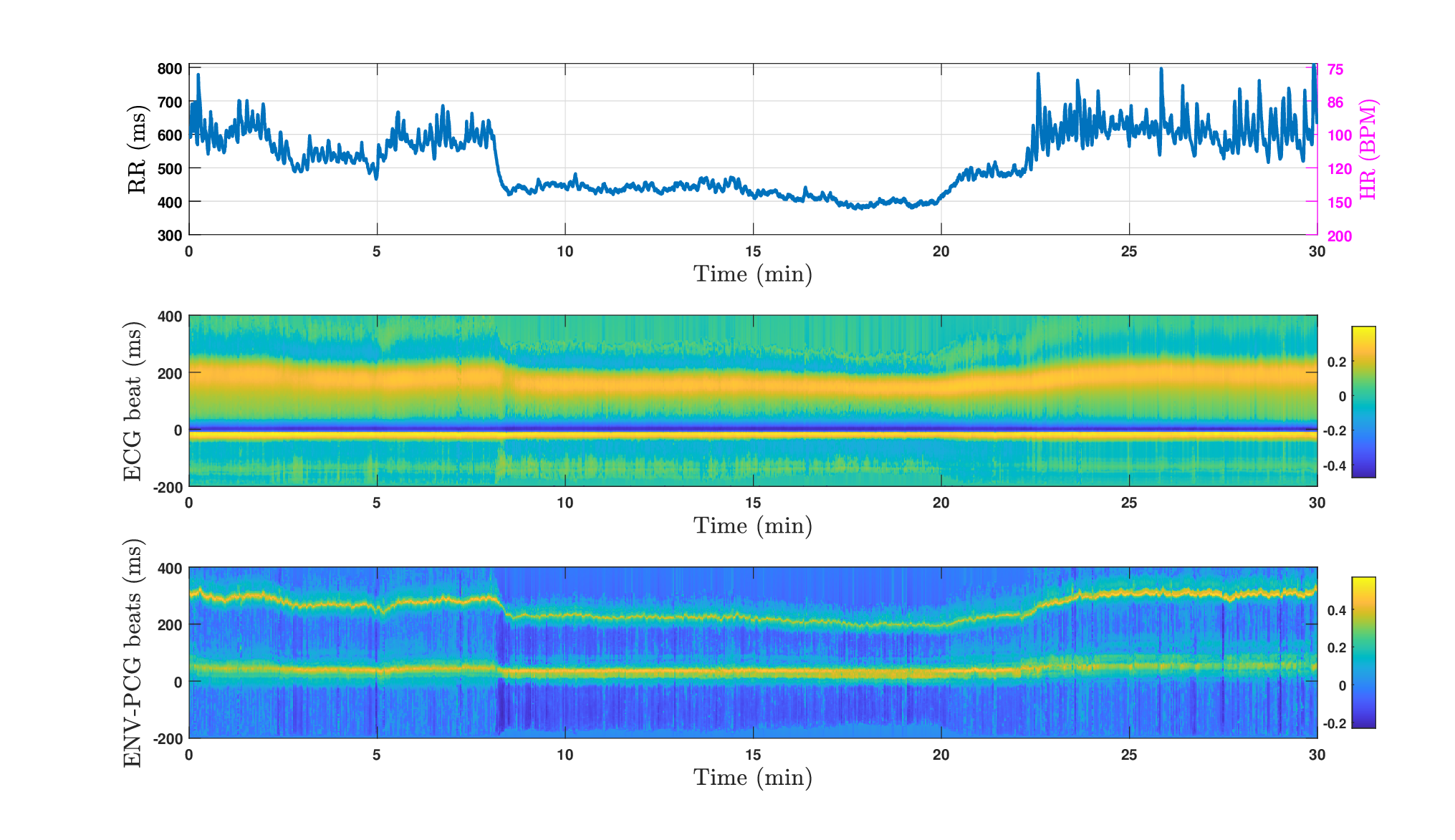}
    \label{fig:waterfall-a}
    \caption{Detailed insights into ECG and PCG changes exercise and recovery responses with the waterfall plot of ECG beats, PCG envelope and RR intervals during a pedaling on a stationary bicycle exercise for record \texttt{ECGPCG0001}. The beats are aligned such that the vertical zero corresponds to the R-peak location of each beat. Notice the changes in the T-wave, S1 and S2 positions (second and third plots) at different heart rates (top plot).}
\label{fig:waterfall}
\end{figure*}

Fig.~\ref{fig:waterfall} shows waterfall plots depicting the extracted ECG beats and PCG envelope beats during a stationary bicycle pedaling recording. The top plot illustrates the RR interval, and the middle plot shows the ECG beat waterfall (a color-coded representation of the ECG beat over time) within a time window of 200\,ms before the R-peak and 400\,ms after the peak for all ECG and PCG beats. In this figure, each vertical color bar corresponds to a single beat. The bottom plot presents the waterfall plot for PCG envelopes within the same time frame. In both waterfall plots, the ECG R-peak is aligned on zero for comparison. The variations in the T-wave width and position, and the displacement of the S1 and S2 components of the PCG demonstrate the cardiac activity dynamics during a stress test. Accordingly, the ECG waterfall plot in Fig.~\ref{fig:waterfall} reveals a prominent yellow band around 150-200\,ms, indicating the T-wave. The T-wave peak and T-wave offset positions in this waterfall suggest a strong correlation between the RR intervals and the QT intervals, as expected. In the PCG waterfall, two distinct yellow strips can be observed, corresponding to the S1 and S2 envelopes. The distance between these two strips represents the systolic interval. As apparent in the waterfall plots, during a stress test, the S2 position adjusts in accordance with the RR interval, while the S1 position remains relatively constant. Furthermore, the S1 peaks occur after the R-peak, as the first yellow strip is located after the zeros position in Y-axes, which corresponds to the R-peak position.

\subsection{Time delay analysis}
To analyze the time delay between extracted ECG and PCG-based intervals, we calculated the cross-correlation function (CCF), to investigate the relationship and coupling between RR intervals and other intervals. To note, CCF has the interesting property that for Gaussian noise it yields the optimal maximum likelihood time-delay estimator~\cite{romagnoli2023model}. We used the normalized CCF as a function of the time delay, obtained by normalizing the cross correlation function between $[-1,1]$, to make the comparison easier. As a result, the CCF assesses the similarity between a time-series RR and lagged versions of other time-series, such as the QT interval, as a function of their time lag. By examining the relationship between these intervals, we can gain valuable insights into the interactions between the electrical and mechanical activities of the heart during stress tests.

\begin{figure*}
    \centering
    \begin{subfigure}{0.3\linewidth}
        \centering
        \includegraphics[width=\columnwidth]{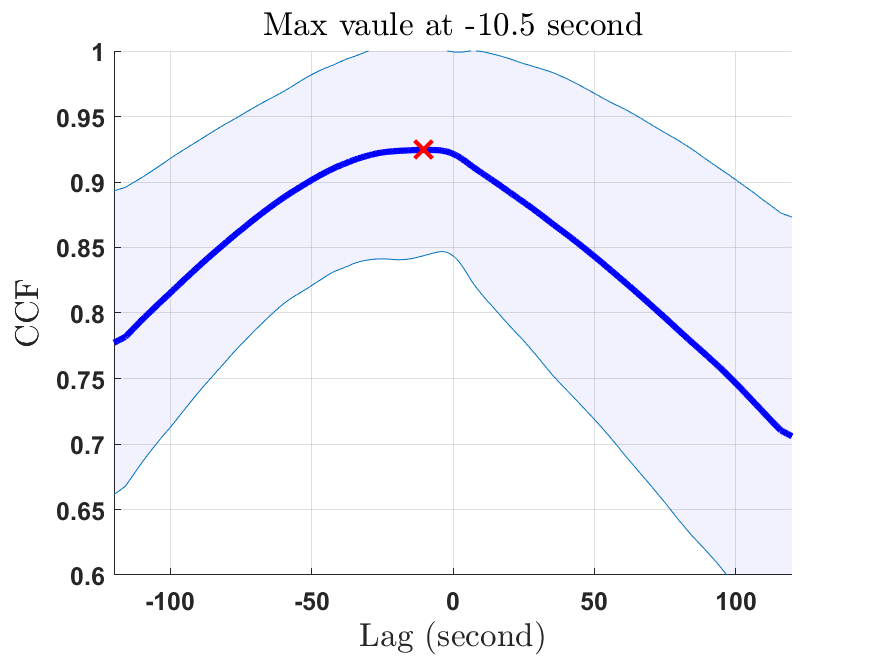}
        \caption{CCF between RR and QT}
        \label{fig:CCFqt}
    \end{subfigure}
    ~
    \begin{subfigure}{0.3\linewidth}
        \centering
        \includegraphics[width=\columnwidth]{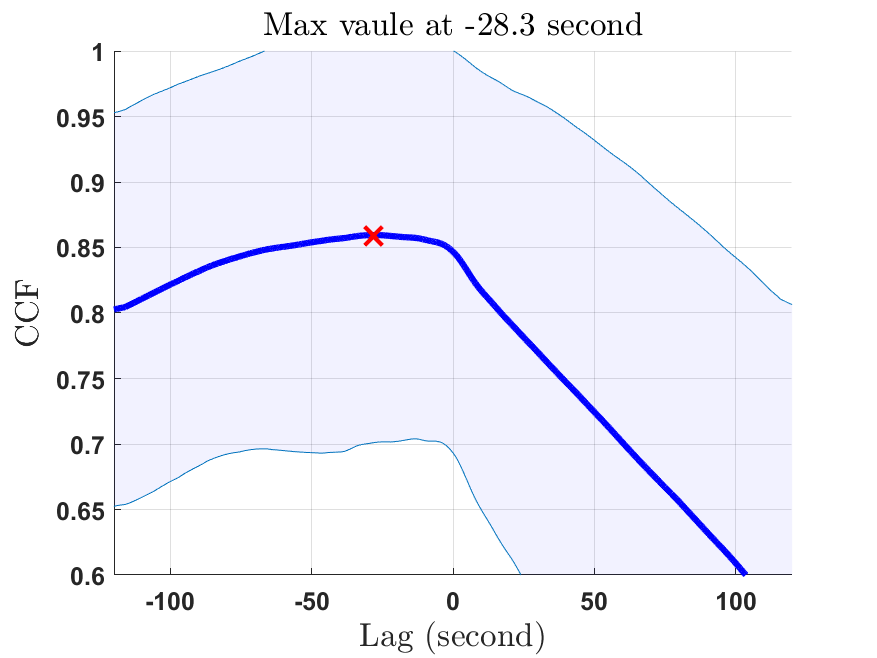}
        \caption{CCF between RR and systolic}
        \label{fig:CCFsys}
    \end{subfigure}
    ~
    \begin{subfigure}{0.3\linewidth}
        \centering
        \includegraphics[width=\columnwidth]{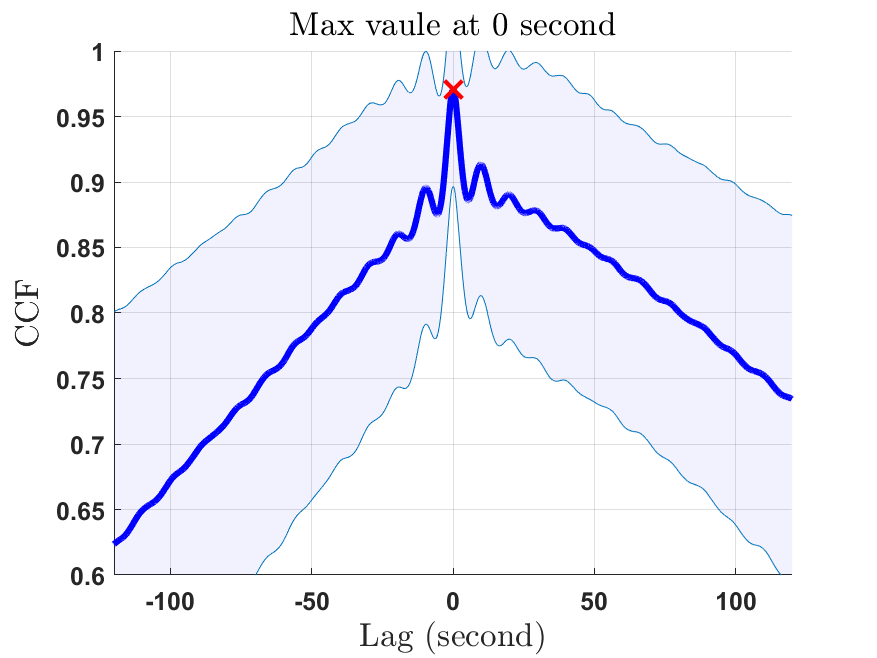}
        \caption{CCF between RR and diastolic}
        \label{fig:CCFdias}
    \end{subfigure}
    \caption{ Cross-correlations functions (mean $\pm$ STD) between RR interval and QT, systolic and diastolic intervals for 23 records.}
\label{fig:CCF}
\end{figure*}

Fig.~\ref{fig:CCF} displays the CCF between RR intervals and the other three intervals for varying time lags across all stress test recordings. The bold blue line in the figure represents the mean, while the shaded area indicates the standard deviation, demonstrating the dispersion of the data around the mean for all stress test recordings. Fig.~\ref{fig:CCF} displays CCF values for lags in the range of -120 to 120 seconds. In addition, the red time sign indicates the maximum value and its specified lag for maximum value has been brought to the title of each plot.

Based on the findings depicted in Fig.~\ref{fig:CCF}, the time delay analysis reveals interesting patterns. The CCF between RR and diastolic intervals peaks at lag zero, indicating simultaneous changes and variations between these intervals. In contrast, the CCF between RR and QT intervals reaches its maximum at -10.5\,s (QT lagging behind RR interval), suggesting that QT variations align more closely with RR changes compared to the systolic intervals, which exhibit their peak correlation at -28.3\,s (systolic lagging behind RR interval). This observation implies that QT interval variations respond more rapidly to RR changes than systolic intervals, highlighting the dynamic relationship between these cardiac parameters. Furthermore, the maximum values of the CCF support the delay analysis. The maximum CCF value for the diastolic interval is 0.97, while it is 0.93 for the QT interval and 0.86 for the systolic interval. These values exhibit an inverse relationship with their lags (a higher lag corresponds to a lower CCF value). This observation suggests that the degree of correlation between the RR intervals and the QT interval decreases as the lag time increases, indicating the dynamic relationship between these cardiac parameters.

Note that, an implicit assumption for using CCF is that the time delay between the RR interval and the corresponding interval (QT interval, in our results) is fixed. We will evaluate this hypothesis vs variable time delays in later sections.

\subsection{Hysteresis effect in response time}
In this section, we initially present visualizations illustrating the hysteresis effect for QT-RR, systolic-RR, and diastolic-RR intervals. Following this, we quantify the relationship between the area-equivalent diameter ($D_a$) of the RR-QT hysteresis loop based on changes in heart rates.

\begin{figure}[tb]
    \includegraphics[trim={2cm 0cm 0cm 0cm},clip,width=0.99\linewidth]{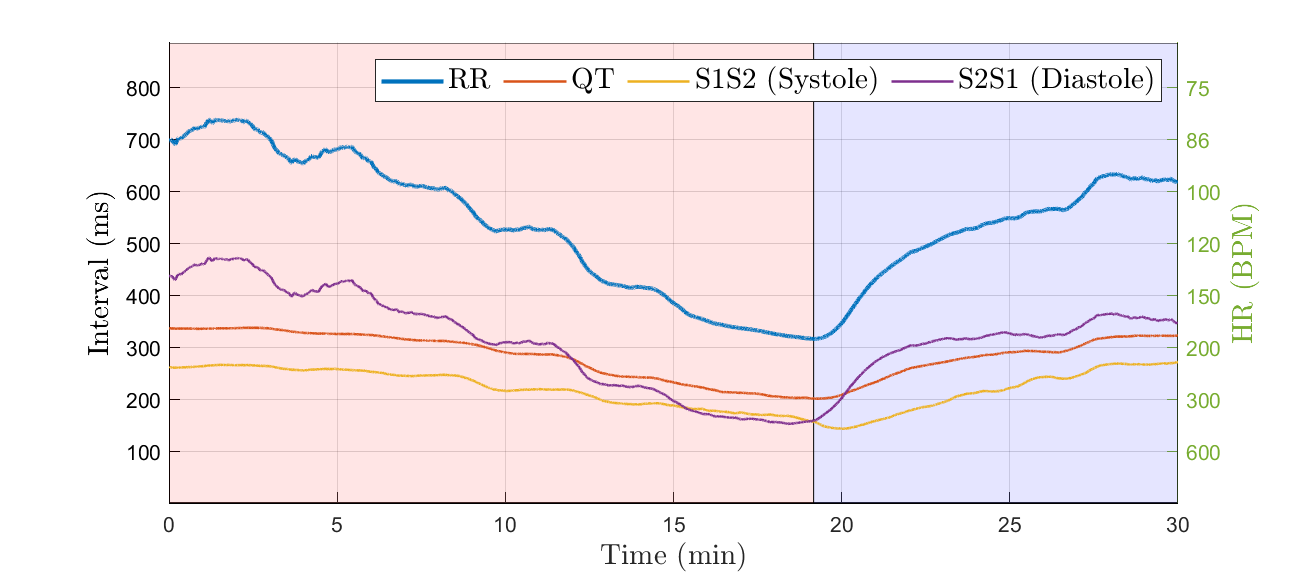}
\caption{Smoothed electromechanical intervals over 1\,min sliding windows, including RR, QT, systolic, and diastolic intervals for record \texttt{ECGPCG0038} (exercise: red; recovery: blue).}
\label{fig:smoothed-intervals}
\end{figure}

To enhance visualization and minimize variability, we applied moving average smoothing with a one-minute duration to all electromechanical intervals. Fig.~\ref{fig:smoothed-intervals} displays the smoothed electromechanical intervals for record \texttt{ECGPCG0038}, following the presentation of the original intervals in Fig.~\ref{fig:Intervals}. moving average operator is a linear phase filter (or constant group delay), ensuring the phase delays between the electromechanical intervals are preserved.

\begin{figure*}[tb]
    \begin{subfigure}[c]{\linewidth}\centering
    \includegraphics[trim={4cm 0cm 4cm 0cm},clip,width=0.98\linewidth]{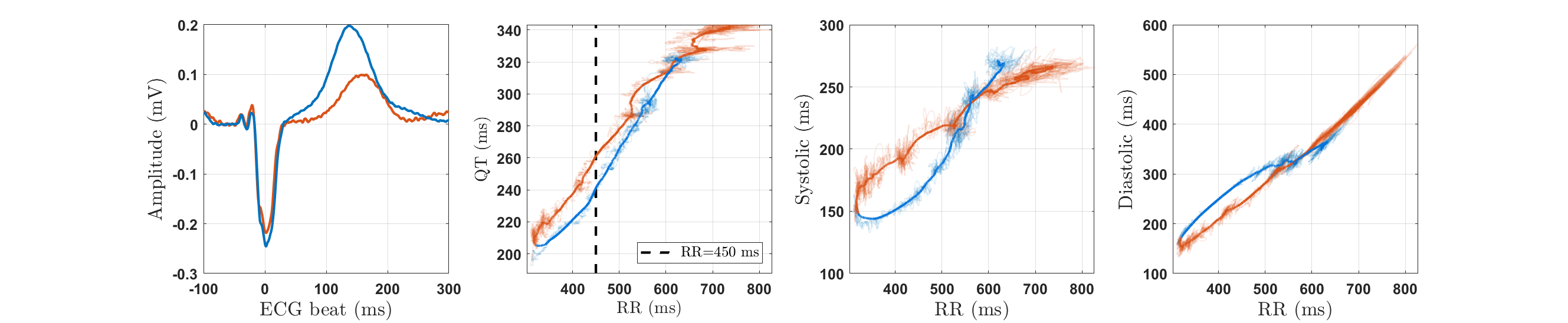}
    \caption{Bruce treadmill stress test for record \texttt{ECGPCG0038}}
    \label{fig:smoothed-state-space-a}
    \end{subfigure}\hfill
    \begin{subfigure}[c]{\linewidth}\centering
    \includegraphics[trim={4cm 0cm 4cm 0cm},clip,width=0.98\linewidth]{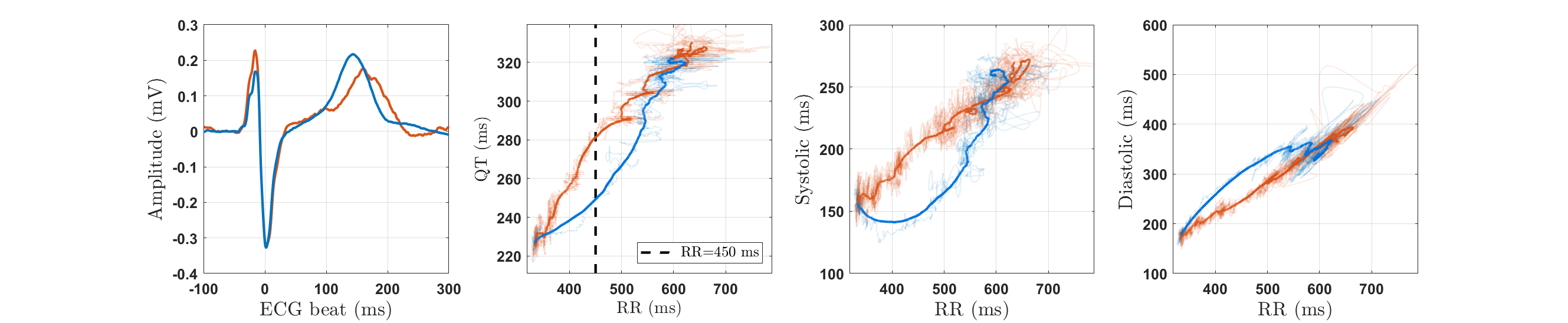}
    \caption{Bruce treadmill stress test for record \texttt{ECGPCG0037}}
    \label{fig:smoothed-state-space-b}
    \end{subfigure}\hfill
    \begin{subfigure}[c]{\linewidth}\centering
    \includegraphics[trim={4cm 0cm 4cm 0cm},clip,width=0.98\linewidth]{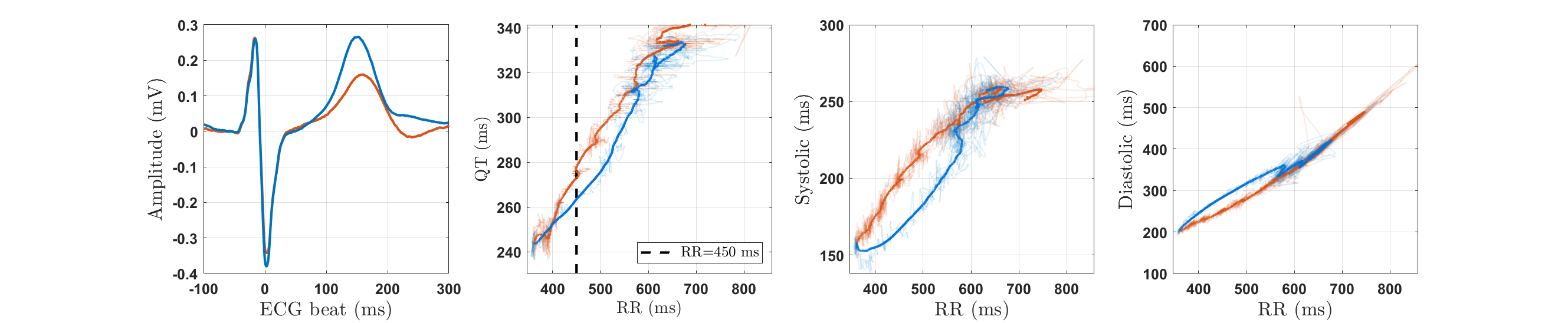}
    \caption{Bicycle stress test for record \texttt{ECGPCG0068}}
    \label{fig:smoothed-state-space-c} 
    \end{subfigure}
\caption{Average ECG beat for exercise and recovery phases at an RR interval of 450\,ms (HR: 133.3\,bpm), along with smoothed interval trajectories in state-space from three stress test recordings. The red paths represent the exercise phase, while the blue paths indicate the recovery phase. For reference, the original unfiltered trajectories are shown with reduced opacity.}
\label{fig:smoothed-state-space}
\end{figure*}

The second to fourth columns in Fig.~\ref{fig:smoothed-state-space} depict the smoothed intervals and the RR interval in state space for three stress test recordings. In these plots, the red path represents the exercise phase, while the blue path signifies the recovery phase. The original unfiltered path is shown in the background with low opacity, providing context for the smoothed trajectories.

In Fig.~\ref{fig:smoothed-state-space}, the first column illustrates the average ECG beat for both the exercise and recovery phases at an RR interval of 450\,ms, as indicated by the dashed line in the QT-RR hysteresis plot. Despite corresponding to the same heart rate, these beats show significant differences in their T-wave morphology during exercise and relation phases. Notably, the T-wave during the recovery phase (blue line) is shifted closer to the QRS complex and has a higher amplitude compared to the T-wave during the exercise phase (red line). This highlights the dynamic changes in cardiac intervals during the stress test, showcasing distinct trajectories observed during the exercise and recovery phases of the heart during stress tests. The clear distinction of the red (exercise) and blue (relaxation) paths indicates the presence of hysteresis in the interval relationships.

We further explore the relationship between heart rate changes in  and the area-equivalent diameter ($D_a$) of the hysteresis loop for QT, systolic, and diastolic intervals. 

\begin{figure*}[tb]
    \centering\includegraphics[trim={0cm 0cm 0cm 0cm},clip,width=0.9\linewidth]{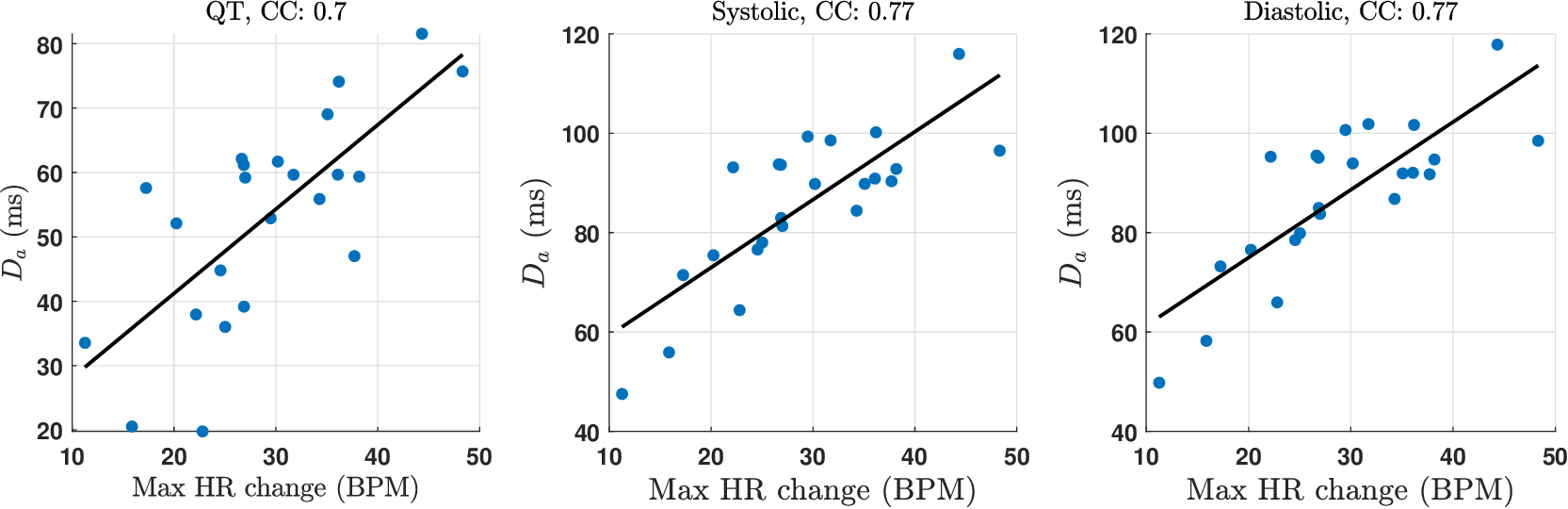}
\caption{The relationship between the equivalent circular area diameter ($D_a$) and the maximum one-minute absolute change in heart rate during recordings. Each dot corresponds to one of the 23 total number of subjects. The high correlation coefficients confirm that higher changes in the heart rate result in more prominent hysteresis effects.}
\label{fig:hysteresis-scatters}
\end{figure*}

Initially, a single $D_a$ value is assigned to each 30-minute recording. Subsequently, the maximum absolute change within sliding one-minute windows is computed for each 30-minute recording in bpm. The scatter plots in Fig.~\ref{fig:hysteresis-scatters} depict the relationship between the maximum absolute change and $D_a$ for QT, systolic, and diastolic hysteresis. The objective is to demonstrate how the characteristic $D_a$ of hysteresis correlates with maximum absolute change in heart rate. Across the three plots, high correlation coefficients (for example 0.75 or above) suggest a robust association between the maximum absolute change and $D_a$, regardless of the subjects. Furthermore, the pattern observed in the scatter plots indicates a similar distribution of $D_a$ for systolic and diastolic hysteresis, aligning with their interdependence, where the sum of both equals the RR intervals.

\subsection{Hysteresis dynamic modeling using machine learning}
The area-equivalent diameter ($D_a$) is a static measure of the cardiac interval coupling. Hypothetically, the inter-dependency between the cardiac intervals is dynamic. In this section, we estimate the QT, systolic, and diastolic intervals from the RR interval using four regression models.

The first model is a memory-less linear model. This model utilizes the instantaneous RR value at a given time $t$ to estimate the \textit{target interval} (QT, systolic, and diastolic intervals) at the same time $t$.

The second model is a memoryless nonlinear neural network, specifically a two-layer fully connected network with a rectified linear uits (ReLU) activation functions. The first layer comprises 50 neurons and implements a clipped ReLU function, while the second layer consists of 25 neurons and uses a leaky ReLU function. In addition, a dropout layer with a dropout rate of 10\% is applied to prevent over-fitting.

The third model is a nonlinear dynamic model, specifically a \textit{long short-term memory} (LSTM) network. This model takes as input the RR intervals from the past minute (600 samples at the 10\,Hz resampled sampling frequency). It features two LSTM layers with 200 and 100 neurons, respectively. The activation functions for this model are leaky ReLUs, and a dropout layer with a rate of 10\% is incorporated. Finally, a linear layer with 25 neurons maps the LSTM outputs into the target intervals.

The fourth model is a \textit{one-dimensional convolutional neural network} (1-D CNN) designed for sequence analysis. Similar to the LSTM model, it uses the RR intervals from the preceding minute (600 samples) as input. This model includes 50 one-dimensional filters of length 5, followed by a one-dimensional max-pooling layer. Subsequently, a fully connected layer with 25 neurons, using a leaky ReLU activation function, maps the CNN filter outputs into the target intervals.

To evaluate the performance of these models, we utilized leave-one-out cross-validation (LOOCV) on 23 selected stress test recordings \cite{karimi2022new}, which involved training the model on 22 recordings and using the left-out recording for testing, with the process repeated until each recording has served as the test set. To maintain consistency in the evaluation period across all models, the first minute of each thirty-minute test recording was excluded from the root mean square error (RMSE) analysis to overcome initial transient effects due to the data preprocessing.

Given that several studies have emphasized the highly individualized relationship between QT and RR intervals, we implemented LOOCV to assess model performance effectively. Additionally, we considered a novel scenario designed to address individual variability patterns. In this scenario, two auxiliary fixed inputs were introduced to the models. Specifically, the mean RR interval and the mean target intervals were calculated for the first minute of each recording, which corresponds to a resting condition at the onset of the stress test with minimal load on the subjects. Therefore, in this scenario, the models were provided with three inputs: RR interval, initial resting RR, and initial resting target values.

RMSE values for the estimation of the three target intervals are presented in Table~\ref{tab:rmse} for both scenarios and all models considered. The results demonstrate that the CNN and LSTM models, when supplemented with two auxiliary initial resting condition inputs, outperform the simpler memory-less models. This suggests that the auxiliary inputs (the cardiovascular baseline at the beginning of the study) are highly effective in capturing individual variability patterns, leading to a significant reduction in RMSE values.
Furthermore, the dynamic LSTM and CNN models achieve the best results, corroborating the dynamic and hysteresis effects underlying the relationship between the RR interval and the QT, systolic, and diastolic intervals, as discussed in the previous section.

\begin{table}[tb]
\centering
\caption{Regression root mean squared errors (RMSE) values in miliseconds for four models in estimating QT, systolic, and diastolic intervals from RR interval, with and without providing initial resting values of the RR and QT intervals.}
\begin{tabular}{|c|c|c|c|}
\hline
\multicolumn{4}{|c|}{\textbf{Models without proving initial resting values}} \\ \hline
\textbf{Model} & \textbf{QT RMSE} & \textbf{Systolic RMSE} & \textbf{Diastolic RMSE} \\ \hline
Linear & 14.8 & 18.7 & 18.8 \\ \hline
NN & 14.5 & 17.4 & 17.4 \\ \hline
LSTM & 13.1 & 17.4 & 17.4 \\ \hline
CNN & 12.8 & 17.6 & 17.6 \\ \hline
\multicolumn{4}{|c|}{\textbf{Models when provided with initial resting values}} \\ \hline
\textbf{Model} & \textbf{QT RMSE} & \textbf{Systolic RMSE} & \textbf{Diastolic RMSE} \\ \hline
Linear & 12.7 & 14.8 & 14.9 \\ \hline
NN & 12.4 & 13.3 & 13.3 \\ \hline
LSTM & 13.5 & \textbf{11.6} & \textbf{11.7} \\ \hline
CNN & \textbf{11.8} & 11.7 & 11.8 \\ \hline
\end{tabular}
\label{tab:rmse}
\end{table}


\section{Discussion}
\label{sec:discussion}
This study provided a comprehensive analysis of the ECG and PCG parameter variations during physical stress test. The observed dynamics of the event intervals extracted from the ECG and PCG, and their correlation with heart rate changes offer valuable insights into the cardiac response to physical stress.
 
The significant variability observed in the S2 to S1 (diastolic) interval depicted in Fig.~\ref{fig:Intervals} indicates that the diastolic period is essential to accommodate HR fluctuations. This observation is consistent with the mechanical response of the heart in varying levels of workload. According to Fig.~\ref{fig:ECGbeats}, as the heart rate increases, the S1 wave power becomes more prominent, while the S2 wave power decreases.

Changes in ECG morphology in Fig.~\ref{fig:ECGbeats} indicate that the ventricular repolarization process exhibits greater sensitivity to changes in heart rate, while ventricular depolarization, as reflected in the QRS complex, remained relatively stable in our data and throughout the range of heart rates examined. The increase in heart rate during exercise was accompanied by a shortening in the duration of the ventricular action potential and the QT interval. As shown in Fig.~\ref{fig:ECGbeats}, this shortening of the QT interval was coupled with a steeper slope of the ST segment in the morphology of the ECG. Previous studies have linked impaired adaptation of the QT interval to changes in heart rate with the risk of cardiac arrhythmias~\cite{pueyo2008dynamic, chevalier2003qt, pathak2005qt}.

The steeper slope of the ST segment at higher heart rates signifies a faster ventricular repolarization process, which is consistent with the increased physiological burden imposed on the heart during exercise. This adaptation allows quicker recovery between beats, facilitating the increased cardiac output required during physical activity.

Furthermore, a minor shift in P-wave morphology and its closer proximity to the R-peak (PR interval shortening) at higher heart rates may indicate changes in atrial conduction times or the influence of increased sympathetic activity during exercise. These findings contribute to our understanding of the dynamic electrical changes in the heart under stress and emphasize the importance of considering heart rate variability when assessing ECG morphology.

The ECG and PCG beat waterfalls provide valuable insights into the dynamic changes in cardiac electrical and mechanical activities during stress test, highlighting the interaction between the various intervals and their responsiveness to the changing heart rate. Furthermore, as seen in Fig.~\ref{fig:waterfall}, the R-wave to S1 segment also decreases with increasing heart rate during exercise. These findings highlight the advantages of multimodal cardiac monitoring, which enables the examination of the interplay between the electrical and mechanical aspects of the heart's performance under stress. 

The cross correlation between RR intervals and other cardiac intervals (QT, systolic, and diastolic) also revealed interesting patterns.

Diastolic intervals showed concurrent changes with RR intervals, peaking at zero lag. QT intervals exhibited a faster response to RR changes, while systolic intervals showed their peak correlation with a larger lag.

In hysteresis analysis, narrower loops corresponded to lower lag, whereas wider loops corresponded to higher lag. Fig.~\ref{fig:smoothed-state-space} demonstrates that the narrowest hysteresis loop in each recording is associated with diastolic-RR hysteresis, while the widest loop is related to the systolic-RR hysteresis. These hysteresis findings align with the results of the time delay analysis.

Several studies have highlighted significant variability in the relationship between QT and RR intervals among healthy young individuals. Fig.~\ref{fig:CCF} shows a wide standard deviation in our subject population, with QT and systolic intervals exhibiting greater variability. This wide standard deviation suggests that the cross-correlation function is highly subject-dependant, with both the peak value and its location varying from one subject to another. The wide standard deviation observed in Fig.~\ref{fig:CCF} aligns and supports previous findings, indicating that electromechanical intervals are uniquely related to the RR interval for each subject~\cite{batchvarov2002individual}.

In future research, understanding the hysteresis phenomenon in cardiovascular responses can be crucial for assessing cardiac performance, especially during stress testing, exercise interventions, and cardiac conditions leading to heart failure. By characterizing hysteresis patterns, clinicians and researchers can gain deeper insights into the dynamic interplay between heart rate and systolic function, facilitating more accurate diagnoses and personalized treatment strategies.

Hysteresis trajectories of the QT-RR and systolic-RR follow similar paths during both exercise and recovery phases, with the exercise (red) path consistently positioned above the recovery (blue) path, as illustrated in Fig.~\ref{fig:smoothed-state-space}. Conversely, hysteresis trajectories for diastolic-RR exhibit opposite directions. This discrepancy stems from the interdependence between systolic and diastolic intervals, wherein their sum equates to the overall intervals, thereby resulting in diastolic-RR trajectories that are expectedly opposite to the systolic-RR patterns.

Hypothetically, the observed differences in T-wave morphology between the exercise and recovery phases, despite maintaining the same heart rate, can be attributed to two physiological factors: autonomic nervous system influence and electrolyte shifts~\cite{billman2013lf}. 
Exercise involves a complex interplay between sympathetic and parasympathetic activity \cite{White2014}. During early exercise, parasympathetic withdrawal dominates up to a heart rate of 100\,bpm, and during continued exercise, sympathetic activity increases, driving the heart rate above 100\,bpm. Recovery is largely characterized by a vagally driven restoration in heart rate to resting levels. This complex interplay between sympathetic and parasympathetic effects may help to explain the reversal of T-amplitude decreases seen in the beginning of exercise; as shown in Fig.~\ref{fig:ecg_beat_superpositions}, the T-amplitude appears to normalize as the heart rate increases above 100\,bpm when sympathetic activity dominates.

This increased sympathetic activity can lead to a more rapid repolarization process, impacting the T-wave morphology. Elevated catecholamine levels, such as adrenaline, during exercise, can shorten the action potential duration, potentially causing the T-wave to appear less pronounced or with a lower amplitude. Conversely, in the recovery phase, there is an increase in parasympathetic (vagal) activity as the body returns to a resting state. This shift prolongs repolarization, making the T-wave more pronounced and with a higher amplitude~\cite{billman2013lf}. The reduced sympathetic tone and increased vagal activity during recovery can lengthen the action potential duration, resulting in the T-wave shifting closer to the QRS complex.
Additionally, intense physical activity can cause shifts in intracellular and extracellular potassium concentrations. High-intensity exercise may lead to transient hyperkalemia (elevated potassium levels in the blood), which can alter the T-wave morphology~\cite{sejersted2000dynamics}. During recovery, as electrolyte balances normalize, the T-wave can exhibit a higher amplitude and different positioning, reflecting the changes in the repolarization process. 

Several studies underscore that the relationship between QT and RR intervals varies greatly among young, healthy individuals, with substantial inter-subject variability observed in the QT-RR relation~\cite{batchvarov2002individual,malik2002relation}. Our findings reveal a significant direct correlation between the hysteresis area ($D_a$) and maximum absolute change in healthy young subjects during stress tests. While our results do not contradict prior findings, they unveil novel nonlinear and dynamic characteristics suggesting a linear relationship between $D_a$ and maximum absolute change among young, healthy individuals. Therefore, $D_a$, and similar methods for measuring hysteresis, can serve as promising marker for hysteresis analysis to evaluate the risk of cardiac arrhythmias during stress tests. machine learning-based models it was demonstrated that the relationships between the heart rate and other ECG/PCG-based biomarkers are effectively modeled by dynamic nonlinear models such as LSTM networks and CNNs. For example, the RMSE values decreased from 14.8\,ms for linear models to 11.7\,ms for LSTM and CNN models when estimating systolic and diastolic intervals from RR intervals, with the inclusion of auxiliary inputs (resting heart rate and QT intervals). Additionally, the inter-subject variability observed in the QT-RR, systolic-RR, and diastolic-RR relationships was more effectively captured by incorporating initial resting condition values as inputs to the machine learning models. For example, this resulted in a reduction of approximately 6\,ms in RMSE values for both LSTM and CNN models in mechanical intervals.

A limitation of our study is the limited population size and demography. In future research, further investigations on larger datasets, more diverse populations with clinical history and across different age groups are warranted to comprehensively assess the utility of hysteresis-based indexes and the value of the interactions between the electrical and mechanical activities of the heart.

\section{Conclusion}
\label{sec:conclusion}
Our exploration of ECG morphology and hysteresis analysis during physical activity has provided valuable insights into cardiac dynamics during stress testing. The sensitivity of ventricular repolarization to heart rate variations, as evidenced by changes in ECG morphology, contrasts with the relative stability of ventricular depolarization. The shortened QT interval at higher heart rates highlights the rapid ventricular repolarization process.
Our results on multimodal analysis of cardiac functions underscores the intricate interplay between electrical and mechanical aspects under stress. The observed shift in the S1 strip with increasing heart rate emphasizes the significance of considering both components in evaluating cardiac performance.
The correlation analysis between RR intervals and other cardiac intervals (QT, systolic, and diastolic) revealed intriguing patterns. Diastolic intervals exhibit synchronous changes with RR intervals, while QT intervals demonstrate a faster response to RR changes. Hysteresis analysis has further elucidated these relationships, with narrower loops associated with diastolic-RR hysteresis and wider loops linked to systolic-RR hysteresis.

In addition, machine learning models confirm the hysteresis phenomena in estimating cardiac intervals using RR interval as input. 
Understanding hysteresis is crucial for diagnosing cardiac conditions, particularly during stress tests. We introduced the effective hysteresis diameter (formulated by $D_a$) as a promising marker for hysteresis analysis, offering insights into the risk of cardiac arrhythmias. Further investigations on larger datasets are essential to validate these novel nonlinear and dynamic characteristics.

Finally, deep learning models successfully estimated the QT, systolic, and diastolic intervals from the RR interval, confirming a nonlinear dynamic relationship between these cardiac intervals. The presence of hysteresis loops highlights the potential of these models to be effectively applied in future research, including stress tests, for estimating cardiac intervals from the RR interval.

In conclusion, our study contributes to the advancement of ECG signal processing methodologies, with the potential to enhance early detection and personalized treatment strategies for heart diseases.

\section*{Acknowledgment}
This research was supported by the American Heart Association Innovative Project Award 23IPA1054351, on ``developing multimodal cardiac biomarkers for cardiovascular-related health assessment.''

\bibliographystyle{IEEEtran}
\bibliography{References}

\begin{thebibliography}{10}
\providecommand{\url}[1]{#1}
\csname url@samestyle\endcsname
\providecommand{\newblock}{\relax}
\providecommand{\bibinfo}[2]{#2}
\providecommand{\BIBentrySTDinterwordspacing}{\spaceskip=0pt\relax}
\providecommand{\BIBentryALTinterwordstretchfactor}{4}
\providecommand{\BIBentryALTinterwordspacing}{\spaceskip=\fontdimen2\font plus
\BIBentryALTinterwordstretchfactor\fontdimen3\font minus \fontdimen4\font\relax}
\providecommand{\BIBforeignlanguage}[2]{{%
\expandafter\ifx\csname l@#1\endcsname\relax
\typeout{** WARNING: IEEEtran.bst: No hyphenation pattern has been}%
\typeout{** loaded for the language `#1'. Using the pattern for}%
\typeout{** the default language instead.}%
\else
\language=\csname l@#1\endcsname
\fi
#2}}
\providecommand{\BIBdecl}{\relax}
\BIBdecl

\bibitem{Oliveira2022}
\BIBentryALTinterwordspacing
J.~Oliveira, F.~Renna, P.~D. Costa, M.~Nogueira, C.~Oliveira, C.~Ferreira, A.~Jorge, S.~Mattos, T.~Hatem, T.~Tavares, A.~Elola, A.~B. Rad, R.~Sameni, G.~D. Clifford, and M.~T. Coimbra, ``{The CirCor DigiScope Dataset: From Murmur Detection to Murmur Classification},'' \emph{IEEE Journal of Biomedical and Health Informatics}, vol.~26, no.~6, p. 2524–2535, Jun. 2022. [Online]. Available: \url{http://dx.doi.org/10.1109/JBHI.2021.3137048}
\BIBentrySTDinterwordspacing

\bibitem{Reyna2023}
\BIBentryALTinterwordspacing
M.~A. Reyna, Y.~Kiarashi, A.~Elola, J.~Oliveira, F.~Renna, A.~Gu, E.~A. Perez~Alday, N.~Sadr, A.~Sharma, J.~Kpodonu, S.~Mattos, M.~T. Coimbra, R.~Sameni, A.~B. Rad, and G.~D. Clifford, ``Heart murmur detection from phonocardiogram recordings: The george b. moody physionet challenge 2022,'' \emph{PLOS Digital Health}, vol.~2, no.~9, p. e0000324, Sep. 2023. [Online]. Available: \url{http://dx.doi.org/10.1371/journal.pdig.0000324}
\BIBentrySTDinterwordspacing

\bibitem{Kazemnejad2024}
\BIBentryALTinterwordspacing
A.~Kazemnejad, S.~Karimi, P.~Gordany, G.~D. Clifford, and R.~Sameni, ``{An open-access simultaneous electrocardiogram and phonocardiogram database},'' \emph{Physiological Measurement}, vol.~45, no.~5, p. 055005, May 2024. [Online]. Available: \url{http://dx.doi.org/10.1088/1361-6579/ad43af}
\BIBentrySTDinterwordspacing

\bibitem{EPHNOGRAMDataset}
\BIBentryALTinterwordspacing
A.~Kazemnejad, P.~Gordany, and R.~Sameni, ``{EPHNOGRAM: A Simultaneous Electrocardiogram and Phonocardiogram Database},'' 2021. [Online]. Available: \url{https://physionet.org/content/ephnogram/1.0.0/}
\BIBentrySTDinterwordspacing

\bibitem{LAUER2006315}
\BIBentryALTinterwordspacing
M.~S. Lauer, C.~E. Pothier, Y.~B. Chernyak, R.~Brunken, M.~Lieber, C.~Apperson-Hansen, and J.~M. Starobin, ``Exercise-induced qt/r-r–interval hysteresis as a predictor of myocardial ischemia,'' \emph{Journal of Electrocardiology}, vol.~39, no.~3, pp. 315--323, 2006. [Online]. Available: \url{https://www.sciencedirect.com/science/article/pii/S0022073605003729}
\BIBentrySTDinterwordspacing

\bibitem{fossa2005dynamic}
A.~A. Fossa, T.~Wisialowski, A.~Magnano, E.~Wolfgang, R.~Winslow, W.~Gorczyca, K.~Crimin, and D.~L. Raunig, ``{Dynamic beat-to-beat modeling of the QT-RR interval relationship: analysis of QT prolongation during alterations of autonomic state versus human ether a-go-go-related gene inhibition},'' \emph{Journal of Pharmacology and Experimental Therapeutics}, vol. 312, no.~1, pp. 1--11, 2005.

\bibitem{STAROBIN2007S91}
\BIBentryALTinterwordspacing
J.~M. Starobin, W.~E. Cascio, A.~H. Goldfarb, V.~Varadarajan, A.~J. Starobin, C.~P. Danford, and T.~A. Johnson, ``Identifying coronary artery flow reduction and ischemia using quasistationary qt/rr-interval hysteresis measurements,'' \emph{Journal of Electrocardiology}, vol.~40, no. 6, Supplement 1, pp. S91--S96, 2007. [Online]. Available: \url{https://www.sciencedirect.com/science/article/pii/S0022073607006486}
\BIBentrySTDinterwordspacing

\bibitem{GravelClinical}
\BIBentryALTinterwordspacing
H.~Gravel, V.~Jacquemet, N.~Dahdah, and D.~Curnier, ``{Clinical applications of QT/RR hysteresis assessment: A systematic review},'' \emph{Annals of Noninvasive Electrocardiology}, vol.~23, no.~1, p. e12514, Oct. 2017. [Online]. Available: \url{https://onlinelibrary.wiley.com/doi/abs/10.1111/anec.12514}
\BIBentrySTDinterwordspacing

\bibitem{studinger2007mechanical}
P.~Studinger, R.~Goldstein, and J.~A. Taylor, ``{Mechanical and neural contributions to hysteresis in the cardiac vagal limb of the arterial baroreflex},'' \emph{The Journal of physiology}, vol. 583, no.~3, pp. 1041--1048, 2007.

\bibitem{liu2013attenuation}
Q.~Liu, B.~P. Yan, C.-M. Yu, Y.-T. Zhang, and C.~C. Poon, ``Attenuation of systolic blood pressure and pulse transit time hysteresis during exercise and recovery in cardiovascular patients,'' \emph{IEEE Transactions on Biomedical Engineering}, vol.~61, no.~2, pp. 346--352, 2013.

\bibitem{willie2011neuromechanical}
C.~K. Willie, P.~N. Ainslie, C.~E. Taylor, H.~Jones, P.~Y. Sin, and Y.-C. Tzeng, ``Neuromechanical features of the cardiac baroreflex after exercise,'' \emph{Hypertension}, vol.~57, no.~5, pp. 927--933, 2011.

\bibitem{martin2022qt}
A.~Mart{\'\i}n-Yebra, L.~S{\"o}rnmo, and P.~Laguna, ``{QT interval adaptation to heart rate changes in atrial fibrillation as a predictor of sudden cardiac death},'' \emph{IEEE Transactions on Biomedical Engineering}, vol.~69, no.~10, pp. 3109--3118, 2022.

\bibitem{gravel2018clinical}
H.~Gravel, V.~Jacquemet, N.~Dahdah, and D.~Curnier, ``{Clinical applications of QT/RR hysteresis assessment: a systematic review},'' \emph{Annals of Noninvasive Electrocardiology}, vol.~23, no.~1, p. e12514, 2018.

\bibitem{perez2023role}
C.~P{\'e}rez, R.~Cebollada, K.~A. Mountris, J.~P. Mart{\'\i}nez, P.~Laguna, and E.~Pueyo, ``{The role of $\beta$-adrenergic stimulation in QT interval adaptation to heart rate during stress test},'' \emph{Plos one}, vol.~18, no.~1, p. e0280901, 2023.

\bibitem{christou2022prolonged}
G.~A. Christou, A.~P. Vlahos, K.~A. Christou, S.~Mantzoukas, C.~A. Drougias, and D.~K. Christodoulou, ``{Prolonged QT interval in athletes: Distinguishing between pathology and physiology},'' \emph{Cardiology}, vol. 147, no. 5-6, pp. 578--586, 2022.

\bibitem{batchvarov2002individual}
V.~Batchvarov and M.~Malik, ``{Individual patterns of QT/RR relationship},'' \emph{Cardiac electrophysiology review}, vol.~6, pp. 282--288, 2002.

\bibitem{malik2002relation}
M.~Malik, P.~F{\"a}rbom, V.~Batchvarov, K.~Hnatkova, and A.~Camm, ``{Relation between QT and RR intervals is highly individual among healthy subjects: implications for heart rate correction of the QT interval},'' \emph{Heart}, vol.~87, no.~3, pp. 220--228, 2002.

\bibitem{bruce1971exercise}
R.~Bruce, ``Exercise testing of patients with coronary artery disease,'' \emph{Ann Clin Res}, vol.~3, pp. 323--332, 1971.

\bibitem{SSJ08}
R.~Sameni, M.~B. Shamsollahi, and C.~Jutten, ``{Model-based Bayesian filtering of cardiac contaminants from biomedical recordings},'' \emph{Physiological Measurement}, vol.~29, no.~5, pp. 595--613, May 2008.

\bibitem{jamshidian2018fetal}
F.~Jamshidian-Tehrani and R.~Sameni, ``{Fetal ECG extraction from time-varying and low-rank noninvasive maternal abdominal recordings},'' \emph{Physiological measurement}, vol.~39, no.~12, p. 125008, 2018.

\bibitem{Pan1985}
J.~Pan and W.~J. Tompkins, ``{A Real-Time QRS Detection Algorithm},'' \emph{Biomedical Engineering, IEEE Transactions on}, vol. BME-32, no.~3, pp. 230--236, 1985.

\bibitem{OSET3.14}
\BIBentryALTinterwordspacing
R.~Sameni, \emph{{The Open-Source Electrophysiological Toolbox (OSET), version 4.0}}, 2006--2024. [Online]. Available: \url{https://github.com/alphanumericslab/OSET.git}
\BIBentrySTDinterwordspacing

\bibitem{karimi2020tractableinf}
S.~Karimi and M.~B. Shamsollahi, ``Tractable inference and observation likelihood evaluation in latent structure influence models,'' \emph{IEEE Transactions on Signal Processing}, vol.~68, pp. 5736--5745, 2020.

\bibitem{karimi2023tractablemle}
S.~{Karimi} and M.~B. {Shamsollahi}, ``Tractable maximum likelihood estimation for latent structure influence models with applications to eeg \& ecog processing,'' \emph{IEEE Transactions on Pattern Analysis and Machine Intelligence}, vol.~45, no.~8, pp. 10\,466--10\,477, 2023.

\bibitem{akram2018analysis}
M.~U. Akram, A.~Shaukat, F.~Hussain, S.~G. Khawaja, W.~H. Butt \emph{et~al.}, ``{Analysis of PCG signals using quality assessment and homomorphic filters for localization and classification of heart sounds},'' \emph{Computer methods and programs in biomedicine}, vol. 164, pp. 143--157, 2018.

\bibitem{cebollada2021mechanisms}
R.~Cebollada, C.~P{\'e}rez, K.~A. Mountris, J.~P. Martinez, P.~Laguna, and E.~Pueyo, ``{Mechanisms underlying QT interval adaptation behind heart rate during stress test},'' in \emph{2021 Computing in Cardiology (CinC)}, vol.~48.\hskip 1em plus 0.5em minus 0.4em\relax IEEE, 2021, pp. 1--4.

\bibitem{montull2023hysteresis}
L.~Montull, {\'O}.~Abenza, R.~Hristovski, and N.~Balagu{\'e}, ``Hysteresis area of psychobiological variables. a new non-invasive biomarker of effort accumulation?'' \emph{Apunts. Educaci{\'o} F{\'\i}sica i Esports}, no. 152, pp. 55--61, 2023.

\bibitem{montull2020hysteresis}
L.~Montull, P.~V{\'a}zquez, R.~Hristovski, and N.~Balagu{\'e}, ``Hysteresis behavior of psychobiological variables during exercise,'' \emph{Psychology of Sport and Exercise}, vol.~48, p. 101647, 2020.

\bibitem{li2005comparison}
M.~Li, D.~Wilkinson, and K.~Patchigolla, ``Comparison of particle size distributions measured using different techniques,'' \emph{Particulate Science and Technology}, vol.~23, no.~3, pp. 265--284, 2005.

\bibitem{leski2002robust}
J.~M. Leski, ``Robust weighted averaging [of biomedical signals],'' \emph{IEEE Transactions on Biomedical Engineering}, vol.~49, no.~8, pp. 796--804, 2002.

\bibitem{romagnoli2023model}
S.~Romagnoli, C.~Per{\'e}z, L.~Burattini, E.~Pueyo, M.~Morettini, A.~Sbrollini, J.~P. Mart{\'\i}nez, and P.~Laguna, ``{Model-based Estimators of QT Series Time Delay in Following Heart-Rate Changes},'' in \emph{2023 45th Annual International Conference of the IEEE Engineering in Medicine \& Biology Society (EMBC)}.\hskip 1em plus 0.5em minus 0.4em\relax IEEE, 2023, pp. 1--4.

\bibitem{karimi2022new}
S.~Karimi and M.~B. Shamsollahi, ``A new post-processing method using latent structure influence models for channel fusion in automatic sleep staging,'' \emph{IEEE Journal of Biomedical and Health Informatics}, vol.~27, no.~3, pp. 1569--1578, 2022.

\bibitem{pueyo2008dynamic}
E.~Pueyo, M.~Malik, and P.~Laguna, ``A dynamic model to characterize beat-to-beat adaptation of repolarization to heart rate changes,'' \emph{Biomedical Signal Processing and Control}, vol.~3, no.~1, pp. 29--43, 2008.

\bibitem{chevalier2003qt}
P.~Chevalier, H.~Burri, P.~Adeleine, G.~Kirkorian, M.~Lopez, A.~Leizorovicz, X.~Andr\'e-fou{\"e}t, P.~Chapon, P.~Rubel, and P.~Touboul, ``{QT dynamicity and sudden death after myocardial infarction: Results of a long-term follow-up study},'' \emph{Journal of cardiovascular electrophysiology}, vol.~14, no.~3, pp. 227--233, 2003.

\bibitem{pathak2005qt}
A.~Pathak, D.~Curnier, J.~Fourcade, J.~Roncalli, P.~K. Stein, P.~Hermant, M.~Bousquet, P.~Massabuau, J.-M. S{\'e}nard, J.-L. Montastruc \emph{et~al.}, ``{QT dynamicity: a prognostic factor for sudden cardiac death in chronic heart failure},'' \emph{European journal of heart failure}, vol.~7, no.~2, pp. 269--275, 2005.

\bibitem{billman2013lf}
G.~E. Billman, ``{The LF/HF ratio does not accurately measure cardiac sympatho-vagal balance},'' \emph{Frontiers in physiology}, vol.~4, p. 45733, 2013.

\bibitem{White2014}
\BIBentryALTinterwordspacing
D.~W. White and P.~B. Raven, ``{Autonomic neural control of heart rate during dynamic exercise: revisited},'' \emph{The Journal of Physiology}, vol. 592, no.~12, p. 2491–2500, May 2014. [Online]. Available: \url{http://dx.doi.org/10.1113/jphysiol.2014.271858}
\BIBentrySTDinterwordspacing

\bibitem{sejersted2000dynamics}
O.~M. Sejersted and G.~Sj{\o}gaard, ``Dynamics and consequences of potassium shifts in skeletal muscle and heart during exercise,'' \emph{Physiological reviews}, vol.~80, no.~4, pp. 1411--1481, 2000.

\end{thebibliography}
\end{document}